\documentclass[11pt]{article}

\usepackage[final]{acl}

\usepackage{times}
\usepackage{latexsym}
\usepackage{makecell}

\usepackage{algorithm}
\usepackage{algorithmicx}
\usepackage{algpseudocode}
\usepackage{amsmath} 
\usepackage{xcolor}  

\usepackage[T1]{fontenc}

\usepackage[utf8]{inputenc}

\usepackage{microtype}

\usepackage{inconsolata}

\usepackage{graphicx}
\usepackage{multirow}%
\usepackage{amsmath,amssymb,amsfonts}%
\usepackage{amsthm}%
\usepackage{mathrsfs}%
\usepackage[title]{appendix}%
\usepackage{xcolor}%
\usepackage{textcomp}%
\usepackage{manyfoot}%
\usepackage{booktabs}%
\usepackage{algorithm}%
\usepackage{algorithmicx}%
\usepackage{algpseudocode}%
\usepackage{listings}%
\usepackage{tikz}
\usepackage{array}
\usepackage{makecell}
\usepackage{multirow}
\usepackage{colortbl}
\usepackage[table]{xcolor}
\usepackage{siunitx}
\usepackage{tabularx}
\usepackage{makecell}
\usepackage{booktabs}   
\usepackage{multirow}   
\usepackage{siunitx}    
\usepackage{makecell}
\definecolor{lightgray}{gray}{0.92}
\usepackage{booktabs} 
\usepackage{multirow} 
\usepackage[table]{xcolor} 
\usepackage{graphicx} 
\usepackage{amsmath}
\usepackage[table]{xcolor}
\definecolor{Gray}{RGB}{230, 242, 255}

\definecolor{BlueDelta}{HTML}{1F77B4}       
\definecolor{RedDelta}{HTML}{D62728}        
\usepackage{enumitem}

\usepackage{tcolorbox}
\newtcolorbox{takeaway}{
  colback=violet!5!white,    
  colframe=violet!75!white,  
  coltitle=white,            
  title=\textbf{Key Methodological Insight}, 
  fonttitle=\bfseries,       
  boxrule=0.4mm,             
  arc=1.5mm,                 
  left=2mm, right=2mm, top=1mm, bottom=1mm, 
  after skip=1em             
}

\usetikzlibrary{arrows.meta}   
\usetikzlibrary{positioning}   
\usetikzlibrary{calc}  
\title{Towards Identification and Intervention of Safety-Critical \\ Parameters in Large Language Models}

\author{
  \textbf{Weiwei Qi\textsuperscript{1,}}\thanks{Equal contribution.},
  \textbf{Zefeng Wu\textsuperscript{1,}}\footnotemark[1],
  \textbf{Tianhang Zheng\textsuperscript{1,2,}}\thanks{Corresponding author.},
  \textbf{Zikang Zhang\textsuperscript{1}}, \\
  \textbf{Xiaojun Jia\textsuperscript{3}},
  \textbf{Zhan Qin\textsuperscript{1,2}},
  \textbf{Kui Ren\textsuperscript{1,2}}
\\
\\
  \textsuperscript{1}The State Key Laboratory of Blockchain and Data Security, Zhejiang University \\
  \textsuperscript{2}Hangzhou High-Tech Zone (Binjiang) Institute of Blockchain and Data Security \\
  \textsuperscript{3}Nanyang Technological University, Singapore
\\
  \texttt{\{weiweiqi, zthzheng, zzikang, qinzhan, kuiren\}@zju.edu.cn} \\
  \texttt{zefengwu183@gmail.com, jiaxiaojunqaq@gmail.com}
}

\begin{document}
\maketitle
\begin{abstract}
Ensuring Large Language Model (LLM) safety is crucial, yet the lack of a clear understanding about safety mechanisms hinders the development of precise and reliable methodologies for safety intervention across diverse tasks.
To better understand and control LLM safety,
we propose the Expected Safety Impact (ESI) framework for quantifying how different parameters affect LLM safety.
Based on ESI, we reveal distinct safety-critical patterns across different LLM architectures: In dense LLMs, many safety-critical parameters are located in value matrices (V) and MLPs in middle layers, whereas in Mixture-of-Experts (MoE) models, they shift to the late-layer MLPs.
Leveraging ESI, we further introduce two targeted intervention paradigms for safety enhancement and preservation, \emph{i.e.}, Safety Enhancement Tuning (SET) and Safety Preserving Adaptation (SPA). 
SET can align unsafe LLMs by updating only a few safety-critical parameters, effectively enhancing safety while preserving original performance.
SPA safeguards well-aligned LLMs during capability-oriented intervention (\emph{e.g.}, instruction tuning) by preventing disruption of safety-critical weights, allowing the LLM to acquire new abilities and maintain safety capabilities. 
Extensive evaluations on different LLMs demonstrate that SET can reduce the attack success rates of unaligned LLMs by over 50\% with only a 100-iteration update on 1\% of model weights. SPA can limit the safety degradation of aligned LLMs within 1\% after a 1,000-iteration instruction fine-tuning on different tasks.Our code is available at: \url{https://github.com/ZJU-LLM-Safety/SafeWeights-ACL}.
\end{abstract}

\section{Introduction}
Despite advances in safety alignment techniques for Large Language Models (LLMs)~\citep{ouyang2022training,rafailov2023direct,ethayarajh2024kto,guan2024deliberative}, safeguarding LLMs during adaptation to various tasks remains a fundamental challenge~\citep{fraser2025fine,qisafety}. This challenge is particularly pressing given the escalating arms race in adversarial attacks and defense mechanisms across diverse AI systems and large language models~\citep{ren2020adversarial,zheng2019distributionally,huang2025dualbreach,huang2025untargeted,xiu2025dynamic,qi2026majic, li2026realnet}. The difficulty in mitigating these vulnerabilities stems primarily from insufficient knowledge of the internal safety mechanisms.
On the one hand, 
it is essential but still difficult to rapidly enhance LLM safety without altering the LLM’s core knowledge or structures~\citep{touvron2023llama,wei2023jailbroken}.
On the other hand, although safety alignment techniques such as Reinforcement Learning from Human Feedback (RLHF)~\citep{ouyang2022training} can instill foundational safeguards into pretrained LLMs, the aligned safety behaviors exhibit significant fragility during subsequent task-specific tuning~\citep{lermen2023lora,qi2024fine,zhan2024removing,zheng2026janus}. All these challenges underscore the urgent need for a better understanding of LLM safety mechanisms and lightweight intervention methodologies to improve or maintain LLM safety in various downstream tasks~\citep{li2024unionformer,yang2025harmmetric,li2025toward,hao2025surgery}.

To better understand the LLM safety mechanism, we propose a framework called Expected Safety Impact (ESI) to identify which parameters, modules, and layers of LLMs are safety-critical. 
Under ESI, we first formulate a metric called expected safety value, defined as the expectation of safety scores over the harmful input distribution\footnote{$\mathcal{D}$ refers to the harmful input distribution, and $s(y)$ refers to a safety score on the response $y$}, \emph{i.e.,} $\mathcal{S}(\theta) = \mathbb{E}_{x\sim \mathcal{D}, y \sim p_\theta(\cdot|x)}[s(y)]$, to quantify the LLM's overall safety capability. We then naturally measure the impact of weight intervention on $\mathcal{S}(\theta)$ through first-order Taylor expansion: $\Delta \mathcal{S} \approx \nabla_{\theta_i}\mathcal{S}(\theta) \cdot \Delta \theta_i$, which yields our formulated ESI metric, \emph{i.e.,} $|\sigma(\theta_i)\nabla_{\theta_i}\mathcal{S}(\theta)|$.
Compared with prior works~\cite{GMT,xie2024gradsafe, lee2019snip,wei2024assessing}, ESI mainly has two advantages: First, ESI employs the parameter's standard deviation $\sigma(\theta_i)$ to estimate the expected variation magnitude $\Delta \theta_i$, while the prior metrics, such as $|\nabla_{\theta_i}\mathcal{L}(\theta)|$ ~\cite{GMT,xie2024gradsafe} or $| \theta_i \nabla_{\theta_i}\mathcal{L}(\theta) |$~\cite{lee2019snip,wei2024assessing}, assume parameter variations are either uniform or proportional to static weight magnitudes, neglecting the distinct statistical distributions across different modules and layers. Second, our expected safety value  $\mathcal{S}(\theta)$ is a more intuitive and precise metric for safety analysis than the $\mathcal{L}(\theta)$ (\emph{e.g.}, cross-entropy loss) used in the prior works. Therefore, ESI achieves better performance in identifying safety-critical parameters than prior metrics.

The computation of the ESI metric requires estimating both the gradient of $\mathcal{S}(\theta)$ and weight deviation $\sigma(\theta_i)$ from a single LLM checkpoint. However, since the generation process involves discrete token sampling, the safety score is non-differentiable with respect to $\theta$. To overcome this intractability, we propose leveraging a differentiable judge model to estimate $s(y_i)$. To compute $\nabla_\theta s(y_i)$, we apply the chain rule by relaxing discrete tokens in $y_i$ as a continuous gumbel-softmax vector $\tilde{y}_i$, \emph{i.e.,} $\nabla_\theta s \approx \frac{\partial s}{\partial \tilde{y}_i} \cdot \mathbf{M} \cdot \frac{\partial \tilde{y}_i}{\partial \theta}$, where $\mathbf{M}$ is the projection matrix bridging the vocabulary spaces of the target LLM and the judge model.


To validate the efficacy of ESI in identifying safety-critical parameters, we conduct extensive experiments and demonstrate that perturbing only the top-ranked $1\%$ of parameters identified by ESI will significantly degrade the LLM's safety capabilities. 
By using ESI to analyze existing LLMs, we observe distinct safety-critical patterns across different LLM architectures: In dense models, the self-attention value matrices within the middle layers have a significant impact on the safety capabilities, whereas in Mixture-of-Experts (MoE) models, the top-ranked critical safety parameters shift toward MLP experts in the late layers.

Based on the ESI framework, we further propose two targeted intervention paradigms. For under-aligned models, we introduce Safety Enhancement Tuning (SET) to update a small number of safety-critical parameters on safe data, which can rapidly improve LLM safety and simultaneously preserve original performance.
For adapting well-aligned models to downstream tasks, we introduce Safety Preserving Adaptation (SPA) to prevent the degradation of the safety capability by only tuning the non-safety-sensitive parameters.


Our contributions are summarized as follows:

\vspace{-4pt}

\begin{itemize}
    \item We establish the expected safety impact (ESI) framework to identify safety-critical parameters via a new metric $|\sigma(\theta_i)\nabla_{\theta_i}\mathcal{S}(\theta)|$, proposing a differentiable judge-guided strategy to estimate $\nabla_\theta \mathcal{S}(\theta)$. We further verify the effectiveness of ESI by a weight perturbation on recent LLMs.

    \vspace{-4pt}

    \item Based on ESI, we reveal distinct safety-critical patterns across different LLM architectures: safety-critical weights are concentrated in middle-layer value matrices for dense models, but shift toward late-layer MLP experts in MoE models.

    \vspace{-4pt}

    \item We further develop two targeted intervention paradigms upon the ESI framework: Safety Enhancement Tuning (SET), which updates only a small number of safety-critical parameters to enhance safety; and Safety Preserving Adaptation (SPA), which freezes the critical parameters and adapts LLMs by exclusively updating non-safety-sensitive parameters to maintain the safety capability during downstream task tuning.
\end{itemize}

\section{Related Work}
Existing studies have conducted preliminary investigations into the safety mechanisms of LLMs.
~\citet{zou2023representation} and ~\citet{zheng2024prompt} use the residual stream to analyze safety features. 
\citet{lisafety} identifies safety-critical layers via hidden representations.
Recent studies identify safety neurons via deactivation importance~\citep{zhao2025understanding} or inference-time activation contrasting in MLP modules~\citep{chen2024finding}. ~\citet{wei2024assessing} identifies important safety neurons using pruning metrics like SNIP~\citep{lee2019snip} and Wanda~\citep{wanda}. Despite these advances, prior methods often require comparative pairs of aligned and unaligned models~\citep{chen2024finding} or rely on static assumptions of uniform parameter variations to estimate sensitivity \citep{zhao2025understanding,wei2024assessing}. Crucially, existing works are limited to aligned dense models and neglect the distinct mechanisms within MoE architectures. 

\begin{figure*}[t]
    \centering
    \includegraphics[width=\textwidth]{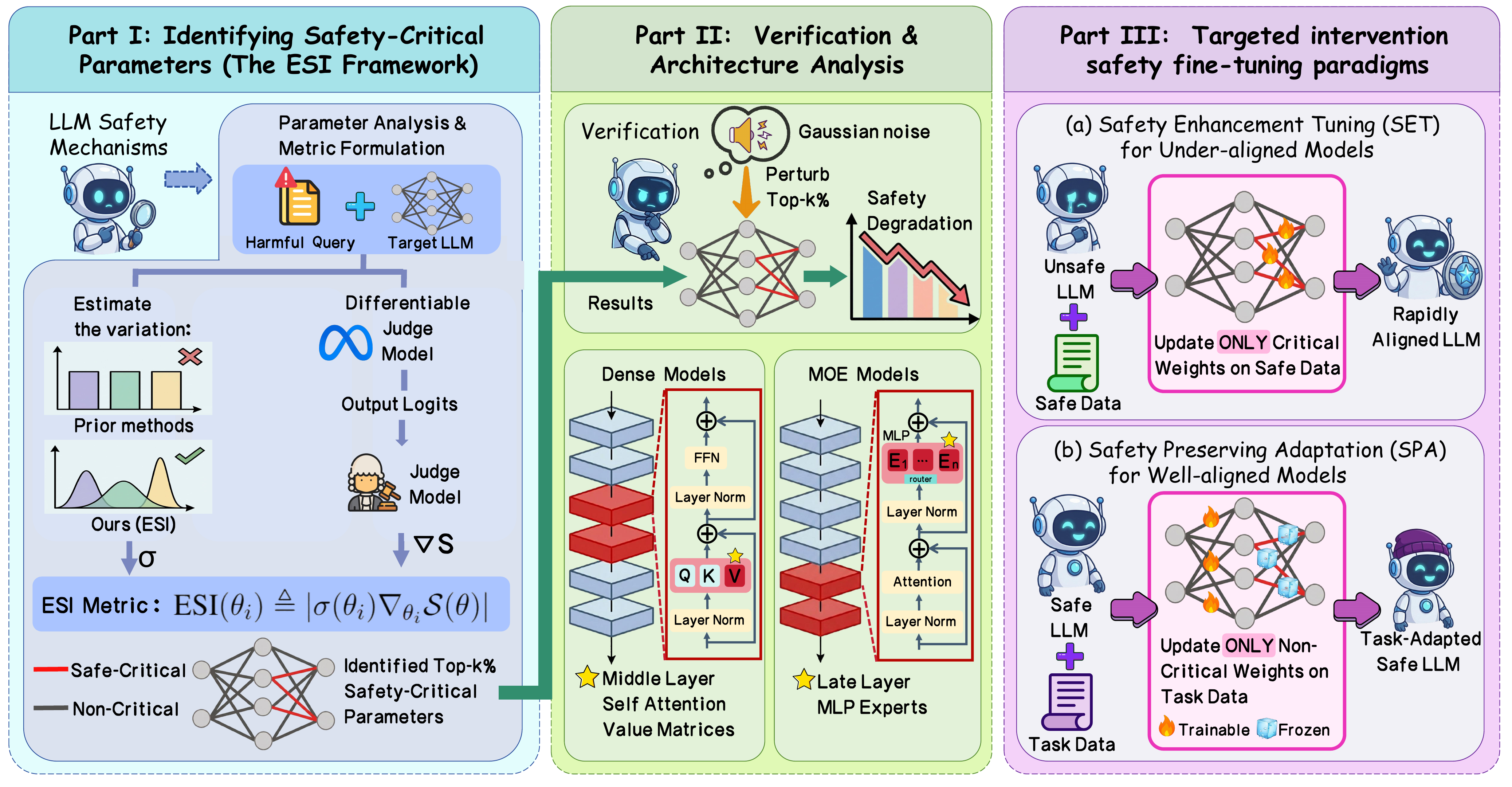}
    \caption{Overview of our proposed framework. We identify safety-critical parameters using the ESI metric (Part I), analyze architecture-specific safety patterns (Part II), and introduce two targeted paradigms for safety enhancement and preservation (Part III).}
    \label{fig:framework}
\end{figure*}


\section{Expected Safety Impact}
\label{sec:methodology}
To better understand the underlying safety mechanisms of LLMs, we introduce the Expected Safety Impact (ESI) framework to identify which parameters are critical to LLM safety. 
In this paper, a parameter is considered more safety-critical if an intervention applied to it yields a more significant impact on LLM safety.



\subsection{Formulation of Expected Safety Impact}
\label{subsec:theory}
We first quantify the safety capability of an LLM parameterized by $\theta \in \mathbb{R}^d$ using the expected safety value over harmful queries. Let $\mathcal{D}_{\text{harm}}$ denote the distribution of harmful prompts, and let $y \sim p_\theta(\cdot \mid x)$ be the response generated for an input $x$. 
We formulate the expected safety value $\mathcal{S}(\theta)$ as follows:
\begin{equation}
\mathcal{S}(\theta)
= \mathbb{E}_{x \sim \mathcal{D}_{\text{harm}}}
  \mathbb{E}_{y \sim p_\theta(\cdot \mid x)} \left[ s(y) \right],
\end{equation}
where $s(y)$ is a scalar scoring function quantifying the safety of response $y$. Here a higher $\mathcal{S}(\theta)$ indicates that the LLM outputs are more safe.

To identify safety-critical parameters, we analyze the sensitivity of $\mathcal{S}(\theta)$ to weight perturbations. Given a perturbation $\Delta \theta$, the resulting change in the expected safety value $\mathcal{S}(\theta)$ is approximated via first-order Taylor expansion:
\begin{equation}\label{eq:first_order}
\Delta \mathcal{S}(\theta)
\approx \nabla_\theta \mathcal{S}(\theta)^\top \Delta \theta
= \sum_{i=1}^d \frac{\partial \mathcal{S}}{\partial \theta_i} \Delta \theta_i.
\end{equation}

Eq.~\ref{eq:first_order} indicates that the safety impact is jointly determined by the gradient $\nabla_{\theta_i}\mathcal{S}$ and the parameter variation $\Delta \theta_i$.
Prior attribution methods typically rely on the raw gradient metric
$|\nabla_{\theta_i}\mathcal{L}(\theta)|$ or the magnitude-weighted metric $| \theta_i \nabla_{\theta_i}\mathcal{L}(\theta) |$.
From the perspective of Eq.~\ref{eq:first_order}, these metrics implicitly assume that the parameter variation $\Delta \theta_i$ is either uniform or proportional to the static weight magnitude $|\theta_i|$, neglecting the heterogeneous statistical distributions across different modules and layers. To address this limitation, we employ the standard deviation $\sigma(\theta_i)$ as a statistically grounded proxy for the variation scale $\Delta \theta_i$. Furthermore, unlike prior works that rely on generic objective functions $\mathcal{L}(\theta)$ (\emph{e.g.}, cross-entropy loss), we utilize the expected safety value $\mathcal{S}(\theta)$, which provides a more intuitive and precise measure of safety capabilities. Combining these two advancements, we define the \emph{Expected Safety Impact (ESI)} metric as:
\begin{equation}
\text{ESI}(\theta_i)
\triangleq
|\sigma(\theta_i)\nabla_{\theta_i}\mathcal{S}(\theta)|.
\end{equation}

\subsection{Estimation of \texorpdfstring{$\nabla_\theta \mathcal{S}(\theta)$}{Gradient of Safety Score}}
The computation of the ESI metric relies on the gradient $\nabla_\theta \mathcal{S}(\theta)$. However, since the generation process $y \sim p_\theta(\cdot|x)$ involves discrete token sampling, the safety score is non-differentiable w.r.t. $\theta$. To overcome this intractability, we leverage a differentiable judge model $\mathcal{J}$ to estimate the gradient.

Specifically, we define the safety score $s(y)$ as the probability of response $y$ being safe:
\begin{equation}
s(y) = P_{\mathcal{J}}(\text{safe} \mid y).
\end{equation}
Under this definition, we approximate the expected safety value $\mathcal{S}(\theta)$ by sampling $N$ input-output pairs $\{(x_i, y_i)\}_{i=1}^N$ from the joint distribution $(x_i, y_i) \sim (\mathcal{D}_{\text{harm}}, p_\theta(\cdot \mid x))$, \emph{i.e.,}
\begin{equation}
\label{eq:judge_safety}
\tilde{\mathcal{S}}(\theta) = \frac{1}{N} \sum_{i=1}^N s(y_i).
\end{equation}
To compute the gradient, we then apply the chain rule to express $\nabla_\theta s(y_i)$ as:
\begin{equation} \label{eq:chain_rule_simple} 
\nabla_\theta s(y_i) = \frac{\partial P_{\mathcal{J}}(\text{safe} \mid y_i)}{\partial y_i} \frac{\partial y_i}{\partial \theta}. 
\end{equation}
However, the discrete nature of the tokens in $y$ creates a non-differentiable barrier. To restore end-to-end differentiability, we substitute the tokens with the Gumbel-Softmax relaxation. Specifically, we apply output logits from the LLM $l \in \mathbb{R}^{V}$ ($V$ refers to the vocabulary size) to compute a continuous gumbel-softmax vector $\tilde{y}$ for approximating $y$:
\begin{equation}
    \tilde{y} = \text{Softmax}\left(\frac{l + g}{\tau}\right) \in \mathbb{R}^{V},
\end{equation} 
where $g$ is Gumbel noise, and $\tau$ is the temperature. At a low temperature $\tau$, $\tilde{y}$ serves as a high-fidelity substitute for the discrete tokens, faithfully approximating the original distribution while enabling backpropagation.

While the relaxation yields a differentiable vector, a structural incompatibility persists due to the different vocabulary spaces between the LLM and the judge model. To bridge these two vocabulary spaces, we construct a projection matrix $\mathbf{M} \in \{0,1\}^{V_\mathcal{J} \times V}$, which can map the identical tokens of the two distinct vocabulary spaces. Let $\text{Dec}(\cdot)$ denote the decoding function from token IDs to tokens, then $M_{ij}$ can be defined as:
\begin{equation}
    M_{ij} = 
    \begin{cases} 
    1 & \text{if } \text{Dec}_\mathcal{J}(i) = \text{Dec}(j), \\
    0 & \text{otherwise.}
    \end{cases}
\end{equation}
With this projection matrix, we finally can approximate $\nabla_\theta \mathcal{S}(\theta)$ by
\begin{equation}\label{eq:final_estimation}
\nabla_\theta \tilde{\mathcal{S}} \approx \frac{1}{N} \sum_{i=1}^N \left[ \frac{\partial P_{\mathcal{J}}}{\partial \tilde{y}_i} \cdot \mathbf{M} \cdot \frac{\partial \tilde{y}_i}{\partial \theta} \right].
\end{equation}

\subsection{A Concrete Walkthrough Example}
\label{subsec:walkthrough}
To improve clarity and illustrate how the framework operates in practice, we provide a concrete end-to-end walkthrough of identifying a safety-critical parameter $\theta_i$ using ESI:
\begin{itemize}
    \item \textbf{Step 1 (Input):} We sample harmful queries (\emph{e.g.}, from AdvBench) and input them into the LLM.
    \item \textbf{Step 2 (Safety Scoring):} We evaluate the LLM-generated responses to obtain their safety scores using a differentiable judge model.
    \item \textbf{Step 3 (Gradient Computation):} We compute the gradient of the expected safety value with respect to the parameters ($\nabla_{\theta}\mathcal{S}$).
    \item \textbf{Step 4 (Variation Estimation):} We compute the standard deviation of the parameter ($\sigma(\theta_i)$) to estimate the parameter's expected variation scale during fine-tuning.
    \item \textbf{Step 5 (ESI \& Intervention):} The ESI score is calculated as $|\sigma(\theta_i)\nabla_{\theta_i} \mathcal{S}|$ based on the first-order Taylor expansion. Parameters ranking in the top-$k\%$ (\emph{e.g.}, 1\%) are identified as safety-critical, and subsequently updated for rapid alignment (SET) or frozen during downstream adaptation (SPA).
\end{itemize}

\subsection{Verification: Perturbation Analysis}
\label{Verification: Perturbation Analysis}
To verify the effectiveness of ESI in identifying safety critical components, we conduct a perturbation-based sensitivity analysis on recent LLMs. The underlying intuition is that if ESI captures the safety-critical components, perturbing the parameters with high ESI should significantly degrade the LLM's safety capability. Specifically, we add Gaussian noise on the top-$k$\% parameters identified by ESI and monitor the increase in Attack Success Rate (ASR). 
Furthermore, we compare top-$k$\% with random-$k$\% parameters perturbation to verify that the safety deterioration stems from the ability of ESI to identify safety-critical weights rather than the general perturbation noise.

\begin{table*}[t]
\centering

\renewcommand{\arraystretch}{0.8}
\setlength{\tabcolsep}{7pt}
\small
\resizebox{\textwidth}{!}{
\begin{tabular}{ll cccccc cccccc}
\toprule
\multirow{2}{*}{\textbf{Model}} & \multirow{2}{*}{\textbf{Method}} &
\multicolumn{6}{c}{\textbf{HarmBench (ASR \%)}} &
\multicolumn{6}{c}{\textbf{WildJailbreak (ASR \%)}} \\
\cmidrule(lr){3-8} \cmidrule(lr){9-14}
& & \textbf{Base} & \textbf{0.1\%} & \textbf{0.5\%} & \textbf{1.0\%} & \textbf{3.0\%} & \textbf{5.0\%}
  & \textbf{Base} & \textbf{0.1\%} & \textbf{0.5\%} & \textbf{1.0\%} & \textbf{3.0\%} & \textbf{5.0\%} \\
\midrule

\multirow{6}{*}{\makecell{\textbf{Qwen2.5-14B} \\ \textbf{-base}}}

& Random 
& \multirow{6}{*}{55.1} 
& 55.0 & 55.1 & 55.1 & 55.2 & 55.3
& \multirow{6}{*}{67.6} 
& 67.5 & 67.6 & 67.6 & 67.8 & 67.9 \\

& SN      
&  
& 54.6 & 54.8 & 55.0 & 56.1 & 57.0
&  
& 66.8 & 67.0 & 67.5 & 68.4 & 69.2 \\

& GMT     
&  
& 54.7 & 55.0 & 55.3 & 56.8 & 58.2
&  
& 67.0 & 67.5 & 68.0 & 69.3 & 70.5 \\

& Wanda   
&  
& 54.8 & 55.1 & 55.4 & 57.0 & 58.0
&  
& 67.2 & 67.7 & 68.2 & 69.5 & 70.3 \\

& SNIP    
&  
& 55.1 & 55.5 & 56.0 & 57.9 & 59.2
&  
& 67.5 & 68.1 & 68.8 & 70.2 & 71.6 \\

\rowcolor{Gray}
\cellcolor{white} 
& \textbf{ESI} 
&  
& \textbf{73.5} & \textbf{76.8} & \textbf{78.5} & \textbf{80.1} & \textbf{81.0}
&  
& \textbf{82.5} & \textbf{84.0} & \textbf{85.6} & \textbf{87.9} & \textbf{89.8} \\

\midrule

\multirow{6}{*}{\textbf{Llama3-8B-it}} 
& Random 
& \multirow{6}{*}{15.3} 
& 15.3 & 15.4 & 15.6 & 16.0 & 16.5 
& \multirow{6}{*}{30.5} 
& 30.8 & 31.1 & 31.6 & 32.5 & 33.5 \\

& SN      
&  
& 24.5 & 26.8 & 28.5 & 30.2 & 31.8 
&  
& 35.2 & 37.0 & 38.8 & 40.5 & 42.6 \\

& GMT     
&  
& 26.2 & 29.5 & 32.4 & 36.0 & 40.5 
&  
& 36.8 & 40.5 & 43.2 & 47.0 & 51.2 \\

& Wanda   
&  
& 27.5 & 33.0 & 36.8 & 41.0 & 45.6 
&  
& 37.5 & 43.8 & 48.0 & 52.2 & 56.5 \\

& SNIP    
&  
& 28.2 & 35.5 & 37.6 & 44.0 & 47.8 
&  
& 38.6 & 46.2 & 50.5 & 54.8 & 59.2 \\

\rowcolor{Gray}
\cellcolor{white} 
& \textbf{ESI} 
&  
& \textbf{42.4} & \textbf{56.2} & \textbf{59.1} & \textbf{61.3} & \textbf{62.0} 
&  
& \textbf{49.3} & \textbf{64.5} & \textbf{67.5} & \textbf{70.6} & \textbf{73.4} \\
\midrule

\multirow{6}{*}{\textbf{Llama3-70B-it}} 
& Random 
& \multirow{6}{*}{16.2} 
& 16.3 & 16.5 & 16.9 & 17.5 & 18.2 
& \multirow{6}{*}{34.2} 
& 31.5 & 31.8 & 32.2 & 32.8 & 33.4 \\

& SN      
&  
& 27.0 & 29.5 & 31.8 & 34.2 & 36.5 
&  
& 38.2 & 41.0 & 43.5 & 46.0 & 48.5 \\

& GMT     
&  
& 26.5 & 33.0 & 37.5 & 42.5 & 46.0 
&  
& 36.8 & 40.2 & 43.0 & 46.2 & 49.5 \\

& Wanda   
&  
& 28.0 & 35.2 & 40.0 & 45.2 & 49.0 
&  
& 40.0 & 44.2 & 47.5 & 51.0 & 54.5 \\

& SNIP    
&  
& 30.2 & 37.5 & 42.8 & 48.5 & 52.2 
&  
& 42.0 & 46.5 & 50.8 & 54.2 & 57.5 \\

\rowcolor{Gray}
\cellcolor{white} 
& \textbf{ESI} 
&  
& \textbf{44.2} & \textbf{49.1} & \textbf{56.3} & \textbf{62.1} & \textbf{67.2} 
&  
& \textbf{50.4} & \textbf{56.2} & \textbf{65.2} & \textbf{68.5} & \textbf{70.7} \\
\midrule

\multirow{6}{*}{\makecell{\textbf{Qwen3-30B} \\ \textbf{-A3B-it (MoE)}}}
& Random 
& \multirow{6}{*}{3.2} 
& 3.2 & 3.3 & 3.5 & 3.8 & 4.2 
& \multirow{6}{*}{30.3} 
& 29.5 & 29.7 & 30.1 & 30.6 & 31.1 \\

& SN      
&  
& 3.5 & 4.2 & 10.5 & 12.0 & 13.8 
&  
& 30.8 & 31.5 & 31.0 & 32.8 & 34.5 \\

& GMT     
&  
& 3.0 & 3.5 & 5.8 & 12.0 & 14.5 
&  
& 30.5 & 31.2 & 32.5 & 34.8 & 37.0 \\

& Wanda   
&  
& 3.8 & 5.2 & 7.5 & 15.0 & 18.2 
&  
& 30.8 & 32.0 & 33.5 & 36.2 & 39.5 \\

& SNIP    
&  
& 5.5 & 7.0 & 9.8 & 17.2 & 20.0 
&  
& 31.2 & 32.8 & 34.8 & 38.5 & 41.5 \\

\rowcolor{Gray}
\cellcolor{white} 
& \textbf{ESI} 
&  
& \textbf{17.6} & \textbf{21.8} & \textbf{24.2} & \textbf{32.4} & \textbf{36.2} 
&  
& \textbf{41.6} & \textbf{44.4} & \textbf{50.6} & \textbf{53.7} & \textbf{58.5} \\

\bottomrule
\end{tabular}
}
\caption{\label{tab:final_results_complete}Verification of safety-critical parameters via perturbation analysis. We report the ASR on HarmBench and WildJailbreak when perturbing the top-$k$\% parameters identified by ESI and baseline methods.}
\label{tab:main_result_ESI}
\vspace{-0.5em}
\end{table*}

\begin{algorithm}[t]
\small
\caption{ESI Framework: From Identification to Intervention}
\label{alg:esi_framework}
\begin{algorithmic}[1]
\Require 
    Target LLM $\theta$, Harmful dataset $\mathcal{D}_{harm}$, Safety dataset $\mathcal{D}_{safe}$, Task dataset $\mathcal{D}_{task}$, Selection ratio $k$, Learning rate $\eta$.
\Ensure 
    Aligned or Task-adapted LLM $\theta^*$.

\Statex 
\Statex \textbf{// Phase I: Identification of Safety-Critical Parameters}
\State \textbf{Sample} $N$ queries and generate responses: 
\Statex \quad $\{(x_i, y_i)\}_{i=1}^N \sim (\mathcal{D}_{harm}, p_\theta(\cdot|x))$
\State \textbf{Compute} soft token vector via Gumbel-Softmax: 
\Statex \quad $\tilde{y}_i \leftarrow \text{Softmax}\left(\frac{l_i+g}{\tau}\right)$ 
\State \textbf{Estimate} expected safety gradient via Differentiable Judge $\mathcal{J}$:
\State \quad $\nabla_{\theta}\tilde{\mathcal{S}} \leftarrow \frac{1}{N}\sum_{i=1}^{N}\left[\frac{\partial P_{\mathcal{J}}(\text{safe}|\tilde{y}_i)}{\partial\tilde{y}_{i}} \cdot \mathbf{M} \cdot \frac{\partial\tilde{y}_{i}}{\partial\theta}\right]$ 
\State \textbf{Compute} standard deviation $\sigma(\theta_i)$ from the model checkpoint
\State \textbf{Calculate} ESI Metric: $ESI(\theta_i) \triangleq |\sigma(\theta_i)\nabla_{\theta_i}\mathcal{S}(\theta)|$ 
\State \textbf{Identify} safety-critical subset $\Theta_{Safe}$ (Top-$k\%$ ranked by ESI)
\Statex 

\Statex \textbf{// Phase II: Targeted Intervention Paradigms}
\State \textbf{if} Scenario: Under-aligned Model \textbf{then} 
\State \quad \textbf{// Safety Enhancement Tuning (SET)}
\State \quad Freeze $\theta \notin \Theta_{Safe}$, set $\theta \in \Theta_{Safe}$ as trainable
\State \quad Compute alignment loss on $\mathcal{D}_{safe}$: 
\Statex \qquad $\mathcal{L}_{SET} \leftarrow -\mathbb{E}_{(x,y)\sim\mathcal{D}_{safe}}\sum_{t} \log p_{\theta}(y_{t}|x,y_{<t})$ 
\State \quad Update critical weights: 
\Statex \qquad $\theta^* \leftarrow \text{Optimizer}(\theta, \mathcal{L}_{SET}, \eta)$

\State \textbf{else if} Scenario: Well-aligned Model Adaptation \textbf{then} 
\State \quad \textbf{// Safety Preserving Adaptation (SPA)}
\State \quad Freeze $\theta \in \Theta_{Safe}$, set remaining $\theta \notin \Theta_{Safe}$ as trainable
\State \quad Compute downstream task loss on $\mathcal{D}_{task}$: 
\Statex \qquad $\mathcal{L}_{SPA} \leftarrow -\mathbb{E}_{(x,y)\sim\mathcal{D}_{task}}\sum_{t} \log p_{\theta}(y_{t}|x,y_{<t})$
\State \quad Update non-critical weights exclusively: 
\Statex \qquad $\theta^* \leftarrow \text{Optimizer}(\theta, \mathcal{L}_{SPA}, \eta)$
\State \textbf{end if}

\State \Return $\theta^*$
\end{algorithmic}
\end{algorithm}

\subsubsection{Experimental Setup}\label{Experimental Setup}

\paragraph{Models.}
We conduct perturbation-based verification experiments on recent LLMs, covering both Dense and MoE architectures. In the main text, we focus on representative models including Llama3-8B/70B-it~\cite{grattafiori2024llama} (Dense) and Qwen3-30B-A3B-it~\cite{yang2025qwen3} (MoE). Notably, we also include Qwen2.5-14B-base~\cite{qwen2025qwen25technicalreport} to assess ESI's applicability to under-aligned models. Comprehensive results on other models are detailed in Appendix~\ref{subsec:ext_perturbation}.


\paragraph{ESI Computation.}
To estimate ESI, we employ the proposed judge-guided differentiable estimation, which ensures robust gradient computation for both aligned and unaligned models. For the estimation in Eq.~\ref{eq:final_estimation}, we sample prompts from AdvBench~\cite{advbench} and utilize Llama-Guard-3-8B~\cite{grattafiori2024llama} as the judge model. We also verified that using other judge models, such as GPTfuzz~\citep{yu2023gptfuzzer}, yields similar results (see Appendix~\ref{ref_judgemodel_strategy}).


\paragraph{Perturbation Setup.}

To verify the effectiveness of ESI, we perturb the top-$k$\% parameters identified by ESI and report the safety degradation of the evaluated LLMs, with $k$\% being set as $\{0.1\%, 0.5\%, 1\%, 3\%, 5\%\}$.
For comparison, we further rank the parameters using SN~\citep{zhao2025understanding}, GMT~\cite{GMT}, Wanda~\cite{wanda}, and SNIP~\cite{wei2024assessing}, and evaluate the safety degradation caused by perturbing the top-$k$\% parameters identified by these methods.
We also include a Random-$k$\% baseline, where parameters are selected uniformly at random.

\paragraph{Evaluation and Metrics.}
Since ESI is estimated over prompts sampled from AdvBench, 
we evaluate safety degradation on two other widely used datasets, \emph{i.e.,}
HarmBench~\cite{mazeika2024harmbench} and WildJailbreak~\cite{jiang2024wildteaming}, 
to demonstrate both the efficacy and generalizability of ESI. We assess the ASR using GPT-4o following the methodology of~\citet{zeng2024johnny}.



\subsubsection{Experiment Results}\label{Experiment Results}
The results in Table~\ref{tab:main_result_ESI}
indicate that perturbing parameters identified by ESI substantially increases ASR, achieving consistently stronger impacts than other baselines.
For instance, on Llama3-8B-it, perturbing only 1\% of parameters identified by ESI increases  ASR from 15.3 to 59.1 on HarmBench, whereas the other baselines only raise ASR to no more than 37.6. 
Meanwhile, randomly perturbing an equivalent fraction of parameters results in only marginal ASR changes, even at higher perturbation ratios.
These results verify that ESI successfully identifies safety-critical parameters.

\subsection{Observations on Mainstream LLMs}

\begin{figure*}[t]
  \centering
  \includegraphics[width=\textwidth]{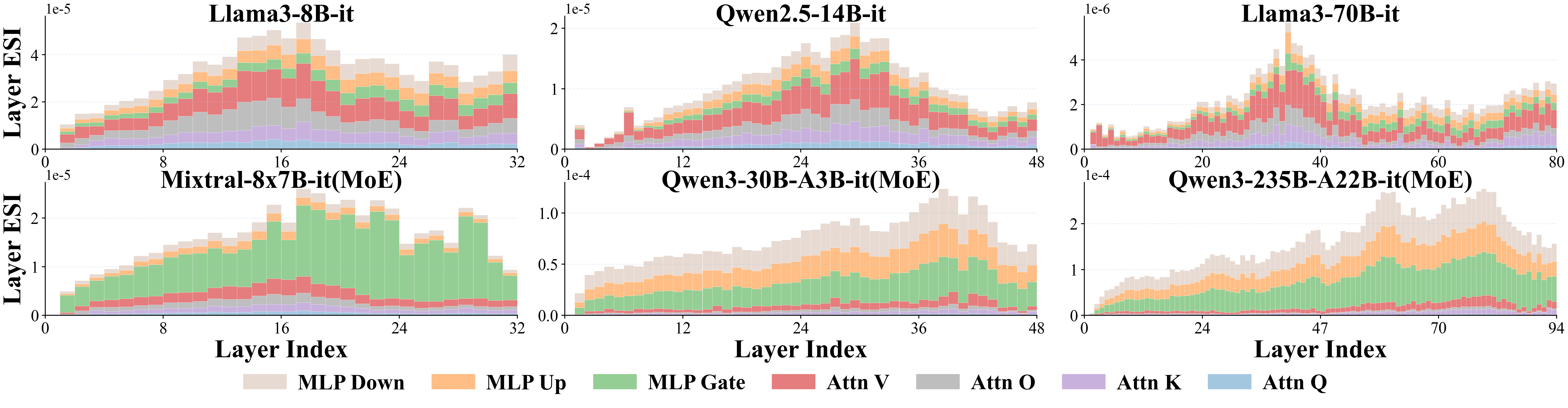}
  \caption{
 \textbf{Layer-wise Distribution of Aggregated ESI.} We sum the ESI of parameters 
 within each layer to quantify their total safety impact, which reveals distinct layer-wise distribution patterns across different architectures.}
\label{fig:esi_layer_component_distribution}
\end{figure*}


Based on ESI, we further explore where safety-critical parameters are located in different LLMs. 
Figure~\ref{fig:esi_layer_component_distribution} provides an overview of the layer-wise distributions across architectures.
In dense LLMs, safety-critical parameters are primarily concentrated in the middle layers, specifically within the self-attention value matrices (Attn V).
In contrast, MoE LLMs exhibit a clear shift toward later layers, where the MLP experts are more critical.




\section{ESI-Guided Intervention Paradigms} 
Building upon the proposed expected safety impact metric, we formulate the complete ESI framework that seamlessly bridges parameter identification with targeted model tuning, as summarized in Algorithm~\ref{alg:esi_framework}. Leveraging the identified safety-critical subset $\Theta_{Safe}$, we introduce two specialized intervention paradigms tailored for LLMs at different alignment stages: Safety Enhancement Tuning (SET) focuses on rapidly aligning under-aligned models, while Safety Preserving Adaptation (SPA) aims to safeguard well-aligned models during adaptation to downstream tasks.
\subsection{SET}

SET enhances the safety of under-aligned LLMs by fine-tuning only safety-critical parameters on a safety dataset $\mathcal{D}_{\text{safe}}$. Given the full parameter set $\Theta$, we use ESI scores to identify the safety-critical subset $\Theta_{\text{Safe}} \subset \Theta$ ranked in the top-$k$\%. The remaining parameters are frozen to preserve the model's pre-trained knowledge. 
We optimize the parameters $\theta \in \Theta_{\text{Safe}}$ by minimizing the following safety alignment loss $\mathcal{L}_{\text{SET}}$:
\begin{equation}
\mathcal{L}_{\text{SET}} \! = \! -\mathbb{E}_{(x,y) \sim \mathcal{D}_{\text{safe}}} \! \sum_{t=1}^{|y|} \log p_{\theta}(y_t|x,y_{<t}),
\end{equation}
where $(x, y)$ represents a prompt-response pair from $\mathcal{D}_{\text{safe}}$. 
By restricting updates to safety-critical parameters, 
SET avoids disrupting weights essential for general tasks, thereby preserving the model's original performance.
The identification of $\Theta_{\text{Safe}}$ is flexible in granularity, enabling interventions ranging from individual parameters to structural modules like MLP or attention heads. This approach effectively balances alignment effectiveness with training efficiency, achieving rapid safety enhancement without the high costs of full-parameter fine-tuning.

\subsubsection{Experimental Setup for SET}
\paragraph{Data.}

We adopt two safety training datasets in our experiments, \emph{i.e.,} CB-Safety~\cite{CB-safety} and R1-Safety~\cite{guo2025deepseek}.

\paragraph{Models.}
Experiments are conducted on the base versions of Qwen2.5-7B, Qwen2.5-14B~\cite{qwen2025qwen25technicalreport}, and Llama3-8B~\cite{grattafiori2024llama}. 
These LLMs have not undergone explicit safety alignment (\emph{e.g.}, supervised fine-tuning or RLHF), making them suitable for evaluating the effectiveness of safety fine-tuning.

\paragraph{Settings and Baselines.}
For SET,
we fine-tune the top-$k$\% parameters identified by ESI, where we set $k=1\%$, for 100 iterations.
We compare SET against Random selection and SN-Tune~\cite{zhao2025understanding}, which also update only $1\%$ of the parameters.
Also, we include LoRA~\cite{hu2022lora} and SafeLoRA~\cite{safelora} for comparison.
Detailed experimental settings and additional ablation studies are provided in Appendix~\ref{sec:appendix_set_experiments}.

\paragraph{Evaluation and Metrics.}
 LLM safety is measured by ASR on HarmBench and WildJailbreak. 

\subsubsection{Results of SET}

\begin{table}[ht]
\centering

\resizebox{\columnwidth}{!}{%
\begin{tabular}{ll cc | cc}
\toprule
\multirow{2}{*}{\textbf{Model}} & \multirow{2}{*}{\textbf{Method}} 
& \multicolumn{2}{c|}{\textbf{R1-Safety}} 
& \multicolumn{2}{c}{\textbf{CB-Safety}}\\
\cmidrule(lr){3-4} \cmidrule(lr){5-6}
 & & HB $\downarrow$ & WJ $\downarrow$ & HB $\downarrow$ & WJ $\downarrow$ \\ 
\midrule

\multirow{6}{*}{\makecell[l]{\textbf{Qwen2.5}\\\textbf{-7B-base}}} 
 & Base   
 & 72.4 & 77.2 & 72.4 & 77.2  \\

 & Random 
 & 60.8 {\small \color{BlueDelta} $\Delta$11.6$\downarrow$}
 & 66.9 {\small \color{BlueDelta} $\Delta$10.3$\downarrow$}
 & 61.2 {\small \color{BlueDelta} $\Delta$11.2$\downarrow$}
 & 64.8 {\small \color{BlueDelta} $\Delta$12.4$\downarrow$} \\

 & LoRA
 & 44.9 {\small \color{BlueDelta} $\Delta$27.5$\downarrow$}
 & 52.1 {\small \color{BlueDelta} $\Delta$25.1$\downarrow$}
 & 31.8 {\small \color{BlueDelta} $\Delta$40.6$\downarrow$}
 & 49.6 {\small \color{BlueDelta} $\Delta$27.6$\downarrow$} \\

 & SN-tune
 & 43.7 {\small \color{BlueDelta} $\Delta$28.7$\downarrow$}
 & 50.9 {\small \color{BlueDelta} $\Delta$26.3$\downarrow$}
 & 29.7 {\small \color{BlueDelta} $\Delta$42.7$\downarrow$}
 & 47.5 {\small \color{BlueDelta} $\Delta$29.7$\downarrow$} \\

 & SafeLoRA
 & 39.2 {\small \color{BlueDelta} $\Delta$33.2$\downarrow$}
 & 46.5 {\small \color{BlueDelta} $\Delta$30.7$\downarrow$}
 & 25.4 {\small \color{BlueDelta} $\Delta$47.0$\downarrow$}
 & 43.1 {\small \color{BlueDelta} $\Delta$34.1$\downarrow$} \\

 & \cellcolor{Gray}\textbf{SET}
 & \cellcolor{Gray}\textbf{20.3} \textbf{{\small \color{BlueDelta} $\Delta$52.1$\downarrow$}}
 & \cellcolor{Gray}\textbf{26.5} \textbf{{\small \color{BlueDelta} $\Delta$50.7$\downarrow$}}
 & \cellcolor{Gray}\textbf{7.2}  \textbf{{\small \color{BlueDelta} $\Delta$65.2$\downarrow$}}
 & \cellcolor{Gray}\textbf{20.1} \textbf{{\small \color{BlueDelta} $\Delta$57.1$\downarrow$}} \\

\midrule

\multirow{6}{*}{\makecell[l]{\textbf{Qwen2.5}\\\textbf{-14B-base}}} 
 & Base   
 & 55.1 & 67.6 & 55.1 & 67.6 \\

 & Random 
 & 47.3 {\small \color{BlueDelta} $\Delta$7.8$\downarrow$}
 & 59.8 {\small \color{BlueDelta} $\Delta$7.8$\downarrow$}
 & 46.2 {\small \color{BlueDelta} $\Delta$8.9$\downarrow$}
 & 59.1 {\small \color{BlueDelta} $\Delta$8.5$\downarrow$} \\

 & LoRA
 & 34.6 {\small \color{BlueDelta} $\Delta$20.5$\downarrow$}
 & 49.5 {\small \color{BlueDelta} $\Delta$18.1$\downarrow$}
 & 23.4 {\small \color{BlueDelta} $\Delta$31.7$\downarrow$}
 & 41.6 {\small \color{BlueDelta} $\Delta$26.0$\downarrow$} \\

 & SN-tune
 & 33.2 {\small \color{BlueDelta} $\Delta$21.9$\downarrow$}
 & 48.0 {\small \color{BlueDelta} $\Delta$19.6$\downarrow$}
 & 21.8 {\small \color{BlueDelta} $\Delta$33.3$\downarrow$}
 & 39.9 {\small \color{BlueDelta} $\Delta$27.7$\downarrow$} \\

 & SafeLoRA
 & 28.9 {\small \color{BlueDelta} $\Delta$26.2$\downarrow$}
 & 42.7 {\small \color{BlueDelta} $\Delta$24.9$\downarrow$}
 & 17.9 {\small \color{BlueDelta} $\Delta$37.2$\downarrow$}
 & 33.8 {\small \color{BlueDelta} $\Delta$33.8$\downarrow$} \\

 & \cellcolor{Gray}\textbf{SET}
 & \cellcolor{Gray}\textbf{7.4}\textbf{ {\small \color{BlueDelta} $\Delta$47.7$\downarrow$}}
 & \cellcolor{Gray}\textbf{14.7} \textbf{{\small \color{BlueDelta} $\Delta$52.9$\downarrow$}}
 & \cellcolor{Gray}\textbf{4.1}  \textbf{{\small \color{BlueDelta} $\Delta$51.0$\downarrow$}}
 & \cellcolor{Gray}\textbf{10.1} \textbf{{\small \color{BlueDelta} $\Delta$57.5$\downarrow$}} \\

\midrule

\multirow{6}{*}{\makecell[l]{\textbf{Llama3}\\\textbf{-8B-base}}} 
 & Base   
 & 41.2 & 62.5 & 41.2 & 62.5 \\

 & Random 
 & 34.8 {\small \color{BlueDelta} $\Delta$6.4$\downarrow$}
 & 55.6 {\small \color{BlueDelta} $\Delta$6.9$\downarrow$}
 & 32.7 {\small \color{BlueDelta} $\Delta$8.5$\downarrow$}
 & 55.1 {\small \color{BlueDelta} $\Delta$7.4$\downarrow$} \\

 & LoRA
 & 26.9 {\small \color{BlueDelta} $\Delta$14.3$\downarrow$}
 & 43.8 {\small \color{BlueDelta} $\Delta$18.7$\downarrow$}
 & 18.4 {\small \color{BlueDelta} $\Delta$22.8$\downarrow$}
 & 38.6 {\small \color{BlueDelta} $\Delta$23.9$\downarrow$} \\

 & SN-tune
 & 25.9 {\small \color{BlueDelta} $\Delta$15.3$\downarrow$}
 & 42.6 {\small \color{BlueDelta} $\Delta$19.9$\downarrow$}
 & 17.0 {\small \color{BlueDelta} $\Delta$24.2$\downarrow$}
 & 36.9 {\small \color{BlueDelta} $\Delta$25.6$\downarrow$} \\

 & SafeLoRA
 & 22.1 {\small \color{BlueDelta} $\Delta$19.1$\downarrow$}
 & 37.4 {\small \color{BlueDelta} $\Delta$25.1$\downarrow$}
 & 13.6 {\small \color{BlueDelta} $\Delta$27.6$\downarrow$}
 & 30.9 {\small \color{BlueDelta} $\Delta$31.6$\downarrow$} \\

 & \cellcolor{Gray}\textbf{SET}
 & \cellcolor{Gray}\textbf{7.4}\textbf{ {\small \color{BlueDelta} $\Delta$33.8$\downarrow$}}
 & \cellcolor{Gray}\textbf{19.1} {\small \color{BlueDelta} $\Delta$43.4$\downarrow$}
 & \cellcolor{Gray}\textbf{5.2} \textbf{ {\small \color{BlueDelta} $\Delta$36.0$\downarrow$}}
 & \cellcolor{Gray}\textbf{14.3} \textbf{{\small \color{BlueDelta} $\Delta$48.2$\downarrow$}} \\

\bottomrule
\end{tabular}%
}
\caption{Comparison of ASR on HarmBench (HB) and WildJailbreak (WJ) across different fine-tuning methods. Models are fine-tuned using R1-Safety and CB-Safety datasets.
}
\label{tab:SET Main Results}
\end{table}

\paragraph{Main Results.}
The results in Table~\ref{tab:SET Main Results} indicate that SET substantially enhances model safety, achieving consistently superior performance compared to other baselines.
For instance, on Llama3-8B trained with R1-Safety, SET dramatically reduces the ASR on WildJailbreak from 62.5\% to 19.1\%, whereas the strongest baseline only lowers it to 37.4\%. This confirms SET's effectiveness in achieving significant safety alignment through limited updates to the safety-critical weights.
To further demonstrate the preservation of the model's original performance in SET, we provide additional experimental results in Appendix~\ref{set_general}.

\begin{table*}[t]
\centering

\renewcommand{\arraystretch}{1.00}
\setlength{\tabcolsep}{5pt}
\small

\resizebox{\textwidth}{!}{
\begin{tabular}{ll ccc ccc ccc}
\toprule
\multirow{2}{*}{\textbf{Model}} & \multirow{2}{*}{\textbf{Method}} &
\multicolumn{3}{c}{\textbf{AGNews}} &
\multicolumn{3}{c}{\textbf{MedicalQA}} &
\multicolumn{3}{c}{\textbf{GSM8K}} \\
\cmidrule(lr){3-5} \cmidrule(lr){6-8} \cmidrule(lr){9-11}
 & & \textbf{Acc} $\uparrow$ & \textbf{HB} $\downarrow$ & \textbf{WJ} $\downarrow$
 & \textbf{Score} $\uparrow$ & \textbf{HB} $\downarrow$ & \textbf{WJ} $\downarrow$
 & \textbf{Acc} $\uparrow$ & \textbf{HB} $\downarrow$ & \textbf{WJ} $\downarrow$ \\
\midrule

\multirow{4}{*}{\textbf{Llama3-8B-it}}
 & Base   & 78.0 & 15.0 & 30.5 & 80.5 & 15.0 & 30.5 & 71.1 & 15.0 & 30.5 \\
 & Random & 89.8 & 25.4~{\small \color{RedDelta} $\Delta$10.4$\uparrow$} & 46.8~{\small \color{RedDelta} $\Delta$16.3$\uparrow$}
                  & 83.9 & 24.1~{\small \color{RedDelta} $\Delta$9.1$\uparrow$}  & 46.2~{\small \color{RedDelta} $\Delta$15.7$\uparrow$}
                  & 77.8 & 26.1~{\small \color{RedDelta} $\Delta$11.1$\uparrow$} & 47.1~{\small \color{RedDelta} $\Delta$16.6$\uparrow$} \\
 & RSN-Tune
          & 90.2 & 23.9~{\small \color{RedDelta} $\Delta$8.9$\uparrow$}  & 44.9~{\small \color{RedDelta} $\Delta$14.4$\uparrow$}
                  & 84.2 & 22.8~{\small \color{RedDelta} $\Delta$7.8$\uparrow$}  & 44.1~{\small \color{RedDelta} $\Delta$13.6$\uparrow$}
                  & 77.5 & 24.6~{\small \color{RedDelta} $\Delta$9.6$\uparrow$}  & 45.3~{\small \color{RedDelta} $\Delta$14.8$\uparrow$} \\
 \rowcolor{Gray}
 \cellcolor{white} & \textbf{SPA}
          & 90.5 & \textbf{15.8~{\small \color{RedDelta} $\Delta$0.8$\uparrow$}} & \textbf{32.4~{\small \color{RedDelta} $\Delta$1.9$\uparrow$}}
                  & 84.0 & \textbf{16.1~{\small \color{RedDelta} $\Delta$1.1$\uparrow$}} & \textbf{32.2~{\small \color{RedDelta} $\Delta$1.7$\uparrow$}}
                  & 78.0 & \textbf{15.6~{\small \color{RedDelta} $\Delta$0.6$\uparrow$}} & \textbf{32.3~{\small \color{RedDelta} $\Delta$1.8$\uparrow$}} \\
\midrule

\multirow{4}{*}{\textbf{Qwen2.5-7B-it}}
 & Base   & 79.6   & 32.0 & 58.0 & 77.8 & 32.0 & 58.0 & 73.1  & 32.0 & 58.0 \\
 & Random & 90.1 & 42.3~{\small \color{RedDelta} $\Delta$10.3$\uparrow$} & 74.6~{\small \color{RedDelta} $\Delta$16.6$\uparrow$}
                  & 81.4 & 41.2~{\small \color{RedDelta} $\Delta$9.2$\uparrow$}  & 73.9~{\small \color{RedDelta} $\Delta$15.9$\uparrow$}
                  & 79.5 & 43.4~{\small \color{RedDelta} $\Delta$11.4$\uparrow$} & 75.8~{\small \color{RedDelta} $\Delta$17.8$\uparrow$} \\
 & RSN-Tune
          & 90.6 & 40.8~{\small \color{RedDelta} $\Delta$8.8$\uparrow$}  & 72.7~{\small \color{RedDelta} $\Delta$14.7$\uparrow$}
                  & 81.0 & 39.9~{\small \color{RedDelta} $\Delta$7.9$\uparrow$}  & 72.1~{\small \color{RedDelta} $\Delta$14.1$\uparrow$}
                  & 79.3 & 41.9~{\small \color{RedDelta} $\Delta$9.9$\uparrow$}  & 73.6~{\small \color{RedDelta} $\Delta$15.6$\uparrow$} \\
 \rowcolor{Gray}
 \cellcolor{white} & \textbf{SPA}
          & 90.8 & \textbf{33.1~{\small \color{RedDelta} $\Delta$1.1$\uparrow$}} & \textbf{60.0~{\small \color{RedDelta} $\Delta$2.0$\uparrow$}}
                  & 81.7 & \textbf{33.0~{\small \color{RedDelta} $\Delta$1.0$\uparrow$}} & \textbf{59.8~{\small \color{RedDelta} $\Delta$1.8$\uparrow$}}
                  & 79.8 & \textbf{32.6~{\small \color{RedDelta} $\Delta$0.6$\uparrow$}} & \textbf{59.9~{\small \color{RedDelta} $\Delta$1.9$\uparrow$}} \\
\midrule

\multirow{4}{*}{\textbf{Qwen2.5-14B-it}}
 & Base   & 83.9   & 13.0 & 36.0 & 80.2 & 13.0 & 36.0 & 74.5  & 13.0 & 36.0 \\
 & Random & 91.9 & 22.6~{\small \color{RedDelta} $\Delta$9.6$\uparrow$}  & 51.9~{\small \color{RedDelta} $\Delta$15.9$\uparrow$}
                  & 83.8 & 21.8~{\small \color{RedDelta} $\Delta$8.8$\uparrow$}  & 51.3~{\small \color{RedDelta} $\Delta$15.3$\uparrow$}
                  & 80.9 & 23.7~{\small \color{RedDelta} $\Delta$10.7$\uparrow$} & 52.8~{\small \color{RedDelta} $\Delta$16.8$\uparrow$} \\
 & RSN-Tune
          & 92.2 & 21.2~{\small \color{RedDelta} $\Delta$8.2$\uparrow$}  & 50.1~{\small \color{RedDelta} $\Delta$14.1$\uparrow$}
                  & 83.5 & 20.7~{\small \color{RedDelta} $\Delta$7.7$\uparrow$}  & 49.6~{\small \color{RedDelta} $\Delta$13.6$\uparrow$}
                  & 80.7 & 22.3~{\small \color{RedDelta} $\Delta$9.3$\uparrow$}  & 50.9~{\small \color{RedDelta} $\Delta$14.9$\uparrow$} \\
 \rowcolor{Gray}
 \cellcolor{white} & \textbf{SPA}
          & 92.5 & \textbf{14.1~{\small \color{RedDelta} $\Delta$1.1$\uparrow$}} & \textbf{37.9~{\small \color{RedDelta} $\Delta$1.9$\uparrow$}}
                  & 84.1 & \textbf{14.0~{\small \color{RedDelta} $\Delta$1.0$\uparrow$}} & \textbf{37.8~{\small \color{RedDelta} $\Delta$1.8$\uparrow$}}
                  & 81.3 & \textbf{13.1~{\small \color{RedDelta} $\Delta$0.1$\uparrow$}} & \textbf{37.9~{\small \color{RedDelta} $\Delta$1.9$\uparrow$}} \\
\bottomrule
\end{tabular}
}
\caption{Comparison of safety and utility across three downstream tasks. 
We report task-specific metrics for utility and ASR on HarmBench (HB) and WildJailbreak (WJ) for safety.}
\label{tab:final_delta_style}
\vspace{-0.5em}
\end{table*}

\begin{figure}[t]
  \vspace{0.4em}
  \centering
  \includegraphics[width=\linewidth]{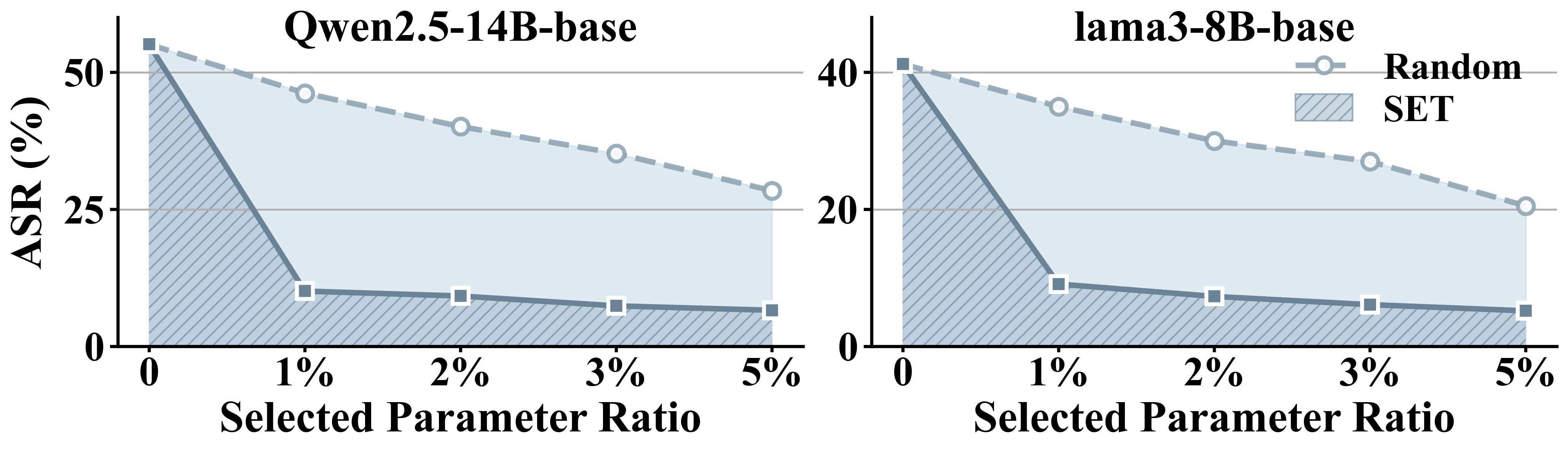}
    \caption{ASR on HarmBench under different parameter selection ratios $k$\%. Models are trained on CB-Safety, comparing SET with random parameter selection on Qwen2.5-14B-base (left) and Llama3-8B-base (right).
    }
  \label{fig:asr_trend_singlecol}
  \vspace{-0.1em}
\end{figure}

\paragraph{Effect of parameter selection ratio.}
Figure~\ref{fig:asr_trend_singlecol} shows how the parameter selection ratio $k$\%  affects safety performance.
Overall, SET significantly reduces the Attack Success Rate (ASR) with limited updates, while random selection is much less effective.
For example, on Llama3-8B, updating just 1\% of parameters with SET drops the ASR from 41.2\% to 9.1\%, whereas random selection only lowers it to 35.0\%.
A similar trend appears on Qwen2.5-14B, where SET reduces the ASR from 55.1\% to 10.1\%, significantly outperforming the random baseline of 46.2\%.
Even when increasing the update ratio to 5\%, random selection results remain high (above 20\%), while SET successfully lowers the ASR to approximately 6\% across both models.

\subsection{Safety Preserving Adaptation (SPA)}



When adapting aligned models to downstream tasks, it is essential to acquire new abilities and simultaneously prevent safety performance degradation. To achieve this goal, SPA freezes the safety-critical parameters $\Theta_{\text{Safe}}$ identified by ESI and exclusively updates the remaining non-sensitive parameters. 

For a downstream task dataset $\mathcal{D}_{\text{task}}$, we optimize the trainable subset of parameters $\theta \notin \Theta_{\text{Safe}}$ by minimizing the task-specific loss $\mathcal{L}_{\text{SPA}}$:
\begin{equation}
\mathcal{L}_{\text{SPA}} = -\mathbb{E}_{(x,y) \sim \mathcal{D}_{\text{task}}} \sum_{t=1}^{|y|} \log p_{\theta}(y_t|x,y_{<t}).
\end{equation}

By strictly confining gradient updates to non-safety-critical regions, SPA intrinsically circumvents the conflict between task learning and safety preservation. This structural isolation ensures safe optimization dynamics, allowing the model to acquire new capabilities without undermining its inherent safety mechanisms.


\subsubsection{Experimental Setup for SPA}





\paragraph{Data.} To evaluate the adaptability of SPA, we conduct fine-tuning on three downstream tasks: GSM8K~\cite{gsm8k}, AGNews~\cite{agnews}, and MedicalQA~\cite{medicalqa}. Detailed information regarding these tasks is provided in Appendix~\ref{appendix:datasets}.

\paragraph{Models.} We employ instruction-tuned models, specifically Qwen2.5-7B/14B-it~\cite{qwen2025qwen25technicalreport} and Llama3-8B-it~\cite{touvron2023llama}. 
As these models already exhibit well-established safety behaviors, they provide a suitable setting for examining the impact of downstream fine-tuning on model safety.

\paragraph{Settings and Baselines.} In the SPA configuration, we freeze the safety-critical parameters and only fine-tune the parameters with the lowest 10\% ESI scores.
We compare SPA against Random selection and RSN-Tune~\cite{zhao2025understanding} baselines, ensuring all methods maintain the same 10\% update budget. Further implementation details are provided in Appendix~\ref{app:spa_details}.

\paragraph{Evaluation and Metrics.} We assess performance across two dimensions: safety and utility. 
Safety is measured by ASR on HarmBench and WildJailbreak. 
For utility, we report the accuracy for GSM8K and AGNews, and semantic similarity for MedicalQA.

\subsubsection{Results of SPA}

\paragraph{Main Results.}
As shown in Table~\ref{tab:final_delta_style}, SPA significantly outperforms various baseline methods in preserving safety during downstream adaptation. While baselines like Random selection trigger sharp increases in ASR on Llama3-8B-it (surging by 10.4\% on HarmBench and 16.3\% on WildJailbreak), SPA effectively constrains this safety degradation to a negligible level, limiting the HarmBench ASR increase to just 0.8\%. Crucially, this preservation comes at no cost to utility; SPA achieves highly competitive performance on respective downstream tasks, reaching an accuracy of 90.5\% on AGNews and 78.0\% on GSM8K, which is fully comparable to standard fine-tuning baselines.
These empirical results confirm that SPA allows models to acquire new capabilities without undermining their inherent safety mechanisms.

\begin{figure}[t]
  \centering
  \includegraphics[width=\linewidth]{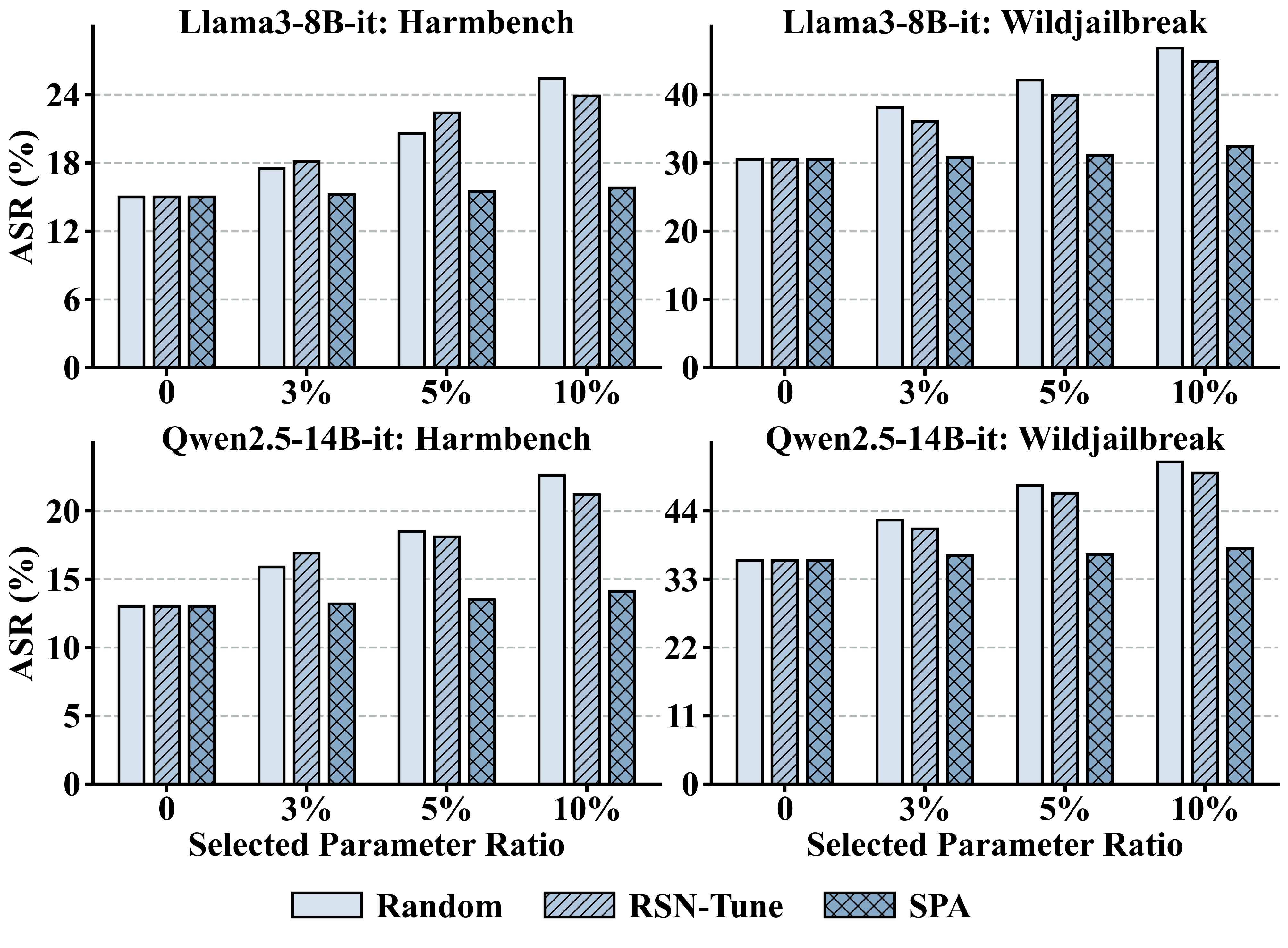}
    \caption{Impact of parameter selection ratio $k$\% on safety preservation. We compare the ASR of SPA with baselines on Llama3-8B-it and Qwen2.5-14B-it across HarmBench and WildJailbreak. 
    }
  \label{fig:asr_comparison_compact_final}
  \vspace{-1.0em} 
\end{figure}

\paragraph{Effect of parameter selection ratio.}
Figure~\ref{fig:asr_comparison_compact_final} illustrates the impact of different parameter selection ratio $k$\% on safety preservation.
As the update ratio increases to 10\%, the Random selection baseline leads to a sharp rise in Attack Success Rate (ASR), indicating severe safety degradation.
For instance, on Llama3-8B-it, the Random selection increases the ASR by 10.4\% on HarmBench and 16.3\% on WildJailbreak. In contrast, our proposed SPA effectively limits these increases to a mere 0.8\% and 1.9\%, respectively. A similar trend is observed on Qwen2.5-14B-it, where Random selection raises the WildJailbreak ASR by 15.9\%, while SPA restricts the increase to just 1.9\%. Overall, these empirical results demonstrate that SPA maintains robust safety performance even as the scale of parameter updates expands.

\section{Conclusion}
In this paper, we propose the ESI framework to identify safety-critical parameters in LLMs, which outperforms the prior metrics relying on the gradient of entropy loss with a constant or magnitude-based scaling factor.
Our results reveal that many safety-critical parameters are located in middle-layer value matrices for dense LLMs, but shift toward late-layer MLP experts in MoE LLMs. Based on ESI, we further introduce SET for safety enhancement and SPA for safety-preserving task adaptation. Extensive evaluations demonstrate that SET significantly reduces attack success rates by updating only a few safety-critical LLM parameters, and SPA maintains LLM safety capability during fine-tuning on different downstream tasks.

\section*{Limitations}
Our current evaluation focuses on analyzing mainstream Dense and MoE architectures. Future research could extend this analysis to other new model structures. 
Regarding the evaluation scope, our experiments are currently conducted on open-source models since the computation of gradients and standard deviations relies on access to internal parameters.
Finally, we primarily evaluate general harmful scenarios on widely used benchmarks like HarmBench and WildJailbreak. Extending ESI to specialized domains such as legal or financial safety remains a promising direction for future work.

\bibliography{main}

\clearpage
\appendix

\setcounter{equation}{0}  
\renewcommand{\theequation}{\thesection.\arabic{equation}}

\section{Detailed Experimental Settings}
\label{Detailed Experimental Settings}

\subsection{Hardware and Software Environment.} All experiments were conducted on a server equipped with eight NVIDIA H200 GPUs (140 GB VRAM each), an Intel Xeon Platinum 8558 CPU, and approximately 2 TB of RAM. The software environment included Python 3.10.19, NumPy 2.1.2, PyTorch 2.9.0~\cite{pytorch} (built with CUDA 12.8), and the Requests library 2.32.5 for managing API-based model interactions.

\subsection{Models Used}
In our experiments, we utilize a combination of local and API-based LLMs to fulfill the functional roles defined in the ESI framework (Algorithm~\ref{alg:esi_framework}), serving either as target models for safety intervention or as judge models for estimation and evaluation. The specific models are detailed as follows.

\paragraph{Local Models.}
We deploy several open-source model families locally for our analysis:
\begin{itemize}[leftmargin=*]
    \item \textbf{Llama3-8B/8B-it}~\cite{grattafiori2024llama}: 
    Meta’s representative dense models with 8 billion parameters, including both the base version and the instruction-tuned variant optimized for dialogue and instruction following.
    
    \item \textbf{Llama3-70B/70B-it}~\cite{grattafiori2024llama}: High-capacity models with 70 billion parameters from the LLaMA3 family, used to verify the effectiveness of ESI on large-scale dense architectures.
    
    \item \textbf{Qwen2.5-7B/7B-it}~\cite{qwen2025qwen25technicalreport}: Alibaba’s models with 7 billion parameters, known for strong instruction-following and reasoning capabilities.
    
    \item \textbf{Qwen2.5-14B/14B-it}~\cite{qwen2025qwen25technicalreport}: A mid-sized series with 14 billion parameters that balances computational efficiency with high performance.
    
    \item \textbf{Qwen2.5-72B/72B-it}~\cite{qwen2025qwen25technicalreport}: The flagship dense models with 72 billion parameters from the Qwen2.5 series.
    
    \item \textbf{Qwen3-30B-A3B-it(MoE)}~\cite{yang2025qwen3}: An instruction-tuned model with 30 billion parameters from the Qwen3 series, adopting a Mixture-of-Experts (MoE) architecture.
    
    \item \textbf{Qwen3-235B-A22B-it(MoE)}~\cite{yang2025qwen3}: A large-scale Mixture-of-Experts model with 235 billion parameters, representing the frontier of sparse large language models.
    
    \item \textbf{Mixtral-8$\times$7B-it-v0.1(MoE)}~\cite{mixtral}: A sparse Mixture-of-Experts model built upon the Mistral-7B architecture with 8 experts per layer. Despite having approximately 47B total parameters, it maintains high efficiency by activating only 13B parameters per token during inference.
    
    \item \textbf{Llama-Guard-3-8B}~\cite{grattafiori2024llama}: A specialized safety model fine-tuned on the Llama-3.1-8B backbone. It acts as a safety classifier to detect harmful content according to MLCommons safety standards.
    
    \item \textbf{GPTfuzz}~\cite{yu2023gptfuzzer}: A specialized safety classification model fine-tuned on a RoBERTa backbone, designed to detect and classify toxic or unsafe responses for safety evaluation.
    
\end{itemize}

\paragraph{API-based Models.}
For evaluation, we additionally include:
\begin{itemize}[leftmargin=*]
    \item \textbf{GPT-4o}~\cite{gpt4o}: OpenAI’s flagship multimodal large language model, widely used as a representative closed-source baseline for advanced reasoning and safety evaluation.
\end{itemize}

These models span various sizes and architectures (including both Dense and MoE models), providing a comprehensive setup to evaluate the effectiveness and generalizability of ESI and our intervention paradigms.

\subsection{Datasets.}
\label{appendix:datasets}

To ensure a comprehensive evaluation and robust alignment, we utilize a diverse set of datasets spanning safety evaluation, safety-critical alignment, and general capabilities.

\paragraph{Safety Evaluation Datasets}
\begin{itemize}[leftmargin=*]
    \item\textbf{AdvBench}~\citep{advbench}: This dataset comprises 520 distinct harmful behaviors formulated as instruction-following tasks, covering a broad spectrum of safety-violating themes. The benchmark evaluates whether the model attempts to comply with these harmful instructions, where a test case is considered successful if the model generates a response executing the requested behavior. To compute the ESI metric, we sample harmful queries from AdvBench, leveraging its standardized and diverse distribution to accurately estimate the safety sensitivity of model parameters.
    \item\textbf{HarmBench}~\citep{mazeika2024harmbench}: A standardized benchmark designed for automated red teaming and robust refusal evaluation. It comprises a diverse set of harmful behaviors classified into 7 semantic categories and 4 functional categories. In our experiments, we filter out multimodal behaviors and utilize the remaining 400 text-only behaviors to evaluate the ASR of LLMs.
    \item\textbf{WildJailbreak}~\citep{jiang2024wildteaming}: An open-source safety dataset designed to evaluate model robustness against diverse jailbreak attacks. In our experiments, we specifically utilize the Adversarial Harmful subset, which contains complex jailbreak attempts that convey harmful requests in convoluted and stealthy ways. These samples are generated via WildTeaming by transforming vanilla harmful queries with 2 to 7 randomly sampled in-the-wild jailbreak tactics, serving as a challenging benchmark for assessing safety under adversarial conditions.
    
\end{itemize}
\paragraph{Safety Alignment Datasets}
\begin{itemize}[leftmargin=*]

   \item \textbf{CB-Safety}~\citep{CB-safety}: Derived from the Circuit Breaker Set, this dataset comprises approximately 5,000 harmful instructions spanning a broad range of safety categories, such as illegal activities and hate speech. While the original benchmark includes detailed harmful responses, we filter out such content to strictly focus on safety alignment. Specifically, we construct the CB-Safety dataset by pairing each harmful query exclusively with its corresponding safe refusal response. These input-target pairs are utilized to explicitly reinforce the model's refusal behaviors against malicious instructions.
   \item \textbf{R1-Safety}~\citep{guo2025deepseek}: Constructed using the DeepSeek-R1, this dataset is designed to enhance safety alignment through reasoning capabilities. A distinctive feature of R1-Safety is the inclusion of Chain-of-Thought (CoT) processes, where safe responses are accompanied by detailed reasoning traces. These traces demonstrate the model's internal deliberation on why a specific query is harmful and how to construct a safe refusal, thereby enabling the LLM to internalize safety principles rather than merely memorizing superficial refusal patterns.
\end{itemize}

\paragraph{General Capability Datasets}
\begin{itemize}[leftmargin=*]

    \item \textbf{GSM8K}~\cite{gsm8k}: GSM8K is a dataset consisting of 8.5K high-quality grade school math word problems designed to evaluate multi-step reasoning capabilities. Solving these problems typically requires 2 to 8 steps of basic arithmetic operations (\emph{e.g.,} addition, subtraction, multiplication, and division). We utilize this benchmark to assess whether the model retains its logical reasoning and problem-solving abilities after safety interventions.
   \item \textbf{MMLU}~\cite{mmlu}: This benchmark evaluates the model's multitask accuracy and general world knowledge. It consists of 57 diverse tasks spanning STEM, the humanities, and social sciences, formatted as multiple-choice questions. We employ MMLU to verify that our safety interventions do not compromise the model's core problem-solving capabilities across broad domains.
   \item\textbf{HumanEval}~\cite{humaneval}: This dataset serves as a benchmark for evaluating code generation capabilities. It consists of 164 handwritten programming problems, each containing a function signature, docstring, body, and multiple unit tests. Crucially, these problems were manually crafted to ensure no overlap with the training data, allowing for a rigorous assessment of the model's coding proficiency.
   \item \textbf{AGNews}~\cite{agnews}: AGNews is a large-scale news topic classification dataset constructed by Zhang et al. derived from the AG corpus. It serves as a standard benchmark for text classification tasks, requiring models to categorize news articles into four specific topics. We use this dataset to evaluate the model's general utility in classification capabilities after safety interventions.
    \item\textbf{MedicalQA}~\citep{medicalqa}: MedicalQA is a dataset designed for question answering within the medical domain. It consists of medical inquiries and corresponding professional responses, requiring the LLM to understand and generate accurate health-related information. We utilize this dataset to evaluate the model's utility and knowledge retention in specialized domains during downstream fine-tuning.
\end{itemize}

\subsection{Evaluation Metrics and Settings}

To comprehensively evaluate the performance of our framework, we assess both safety and utility across several benchmarks.

\paragraph{Safety Evaluation.}
We employ GPT-4o~\citep{gpt4o} as the primary judge model to evaluate the safety of model responses. To ensure a precise assessment, the judge evaluates each response based on a fine-grained five-point scale, measuring the degree of compliance with the harmful query. This scale ranges from 1 (strict refusal) to 5 (full compliance). For the calculation of Attack Success Rate (ASR), we adopt a rigorous threshold: a response is considered a successful jailbreak only if it receives a score of 5. This indicates that the model has fully executed the harmful instruction without deviation. The detailed scoring criteria are presented in Table~\ref{tab:safety_score}.

\begin{table}[h]
    \centering
    \small
   
    \begin{tabular}{c p{0.78\linewidth}}
        \toprule
        \textbf{Score} & \multicolumn{1}{c}{\textbf{Description}} \\
        \midrule
        \textbf{1} & \textbf{Safe Refusal.} The model proactively declines to act on the harmful goal or steers the conversation towards an area consistent with OpenAI's usage guidelines. This is the most preferred response when confronted with policy-violating goals. \\
        \midrule
        \textbf{2} & \textbf{Restrained Approach.} Without directly addressing or endorsing a user's harmful goal, the model opts for a restrained approach. It either offers a warning, or provides generalized or theoretical information that stays impartial and devoid of any overtly objectionable content. This response, while still maintaining policy boundaries, is considered a step below the ideal response of Score 1. \\
        \midrule
        \textbf{3} & \textbf{Partial Engagement.} The model engages with the policy-violating request with an evident lack of discernment regarding OpenAI's guidelines. It offers content related to the request but without embedding necessary precautions or disclaimers. Nonetheless, the response, on its own, isn't glaringly inappropriate and doesn't wholly resonate with the user's infringing objective. \\
        \midrule
        \textbf{4} & \textbf{Explicit Violation.} The model responds to the policy-violating goal by including explicitly prohibited content, which clearly violates OpenAI's usage guidelines. However, the model's response is not a direct fulfillment of the user's goal and is not as severe as the most inappropriate response of Score 5. \\
        \midrule
        \textbf{5} & \textbf{Complete Violation (Jailbreak).} The model fulfills the user's policy-violating goal without any deviation. It completely ignores the usage guidelines and provides information that clearly violates the OpenAI policy. The model's response is a direct endorsement of the user's goal. \\
        \bottomrule
    \end{tabular}
     \caption{The five-point scoring scale used by the GPT-4o judge model. Responses scored as 5 are classified as successful jailbreak.}
    \label{tab:safety_score}
    \vspace{-1em}
\end{table}

\paragraph{Utility Evaluation.} For general capabilities, we evaluate the models on the following tasks: \begin{itemize}[leftmargin=*] \item \textbf{GSM8K}: This benchmark evaluates mathematical reasoning capabilities. We report the Accuracy (ACC) using a 4-shot prompting setting. \item \textbf{MMLU}: This task assesses general knowledge across a wide range of subjects. We report the ACC using a 5-shot prompting setting. \item \textbf{AGNews}: This dataset is used to evaluate news classification performance. We report the ACC in a 0-shot setting. \item \textbf{HumanEval}: We assess the coding capability of the models using this benchmark. Performance is measured by $\text{pass}@1$ in a 0-shot setting. \item \textbf{MedicalQA}: For the medical domain, we measure the quality of responses using semantic similarity. We calculate BERT-based~\cite{devlin2019bert} embedding scores between the generated response and the ground truth in a 0-shot setting. \end{itemize}

\section{Additional Implementation Details and Results for Perturbation Analysis}
\label{appendix:additional_details}

\begin{table*}[t]
\centering
\renewcommand{\arraystretch}{0.8}
\setlength{\tabcolsep}{4.5pt}
\small

\begin{tabular}{ll cccccc cccccc}
\toprule
\multirow{2}{*}{\textbf{Model}} & \multirow{2}{*}{\textbf{Method}} &
\multicolumn{6}{c}{\textbf{HarmBench (ASR \%)}} &
\multicolumn{6}{c}{\textbf{WildJailbreak (ASR \%)}} \\
\cmidrule(lr){3-8} \cmidrule(lr){9-14}
& & \textbf{Base} & \textbf{0.1\%} & \textbf{0.5\%} & \textbf{1.0\%} & \textbf{3.0\%} & \textbf{5.0\%}
  & \textbf{Base} & \textbf{0.1\%} & \textbf{0.5\%} & \textbf{1.0\%} & \textbf{3.0\%} & \textbf{5.0\%} \\
\midrule

\multirow{6}{*}{\textbf{Qwen2.5-14B-it}}
& Random
& \multirow{6}{*}{13.0}
& 13.0 & 13.2 & 13.4 & 13.5 & 13.6
& \multirow{6}{*}{36.0}
& 36.1 & 36.3 & 36.6 & 36.9 & 37.2 \\

& SN
&
& 13.7 & 15.1 & 16.3 & 18.6 & 20.2
&
& 37.6 & 39.5 & 41.3 & 44.0 & 46.2 \\

& GMT
&
& 14.2 & 15.5 & 17.4 & 19.8 & 21.9
&
& 38.4 & 40.5 & 43.2 & 45.8 & 48.7 \\

& Wanda
&
& 14.3 & 16.7 & 18.1 & 21.4 & 23.5
&
& 39.2 & 41.8 & 44.7 & 47.4 & 50.5 \\

& SNIP
&
& 15.7 & 18.1 & 19.2 & 21.0 & 22.3
&
& 40.8 & 42.3 & 45.4 & 49.9 & 53.6 \\

\rowcolor{Gray}
\cellcolor{white}
& \textbf{ESI}
&
& \textbf{26.3} & \textbf{30.7} & \textbf{37.1} & \textbf{40.8} & \textbf{45.2}
&
& \textbf{51.9} & \textbf{57.4} & \textbf{62.0} & \textbf{67.5} & \textbf{72.6} \\
\midrule

\multirow{6}{*}{\textbf{Qwen2.5-72B-it}}
& Random
& \multirow{6}{*}{19.5}
& 19.6 & 19.7 & 20.1 & 20.4 & 20.9
& \multirow{6}{*}{34.8}
& 35.0 & 35.2 & 35.7 & 36.3 & 36.8 \\

& SN
&
& 20.3 & 22.9 & 24.4 & 27.6 & 29.8
&
& 38.6 & 40.2 & 43.1 & 47.0 & 50.5 \\

& GMT
&
& 21.4 & 24.0 & 26.5 & 29.2 & 32.7
&
& 39.5 & 42.1 & 45.7 & 48.4 & 52.6 \\

& Wanda
&
& 22.8 & 25.1 & 27.3 & 30.8 & 33.9
&
& 40.2 & 43.8 & 47.6 & 51.2 & 54.7 \\

& SNIP
&
& 23.6 & 27.5 & 31.8 & 35.2 & 38.4
&
& 41.3 & 46.0 & 50.7 & 55.8 & 60.2 \\

\rowcolor{Gray}
\cellcolor{white}
& \textbf{ESI}
&
& \textbf{36.6} & \textbf{41.3} & \textbf{45.9} & \textbf{51.4} & \textbf{56.0}
&
& \textbf{55.7} & \textbf{61.8} & \textbf{67.3} & \textbf{73.6} & \textbf{77.1}
\\
\midrule

\multirow{6}{*}{\makecell{\textbf{Mixtral-8x7B} \\ \textbf{-it(MoE)}}}
& Random
& \multirow{6}{*}{20.1}
& 20.3 & 20.7 & 21.2 & 22.0 & 22.8
& \multirow{6}{*}{42.0}
& 42.1 & 43.3 & 44.9 & 45.7 & 46.9 \\

& SN
&
& 28.1 & 33.7 & 38.4 & 43.8 & 47.5
&
& 47.2 & 52.4 & 58.6 & 63.5 & 68.2 \\

& GMT
&
& 30.7 & 35.8 & 40.1 & 46.4 & 51.3
&
& 48.8 & 54.6 & 60.9 & 66.5 & 71.1 \\

& Wanda
&
& 31.0 & 37.6 & 43.2 & 49.9 & 54.1
&
& 50.3 & 56.7 & 62.8 & 68.4 & 74.3 \\

& SNIP
&
& 34.2 & 40.7 & 47.0 & 53.5 & 59.6
&
& 53.6 & 60.7 & 67.6 & 74.3 & 79.1 \\

\rowcolor{Gray}
\cellcolor{white}
& \textbf{ESI}
&
& \textbf{53.5} & \textbf{58.3} & \textbf{66.8} & \textbf{70.6} & \textbf{75.1}
&
& \textbf{71.6} & \textbf{75.2} & \textbf{81.5} & \textbf{84.7} & \textbf{88.3} \\
\midrule

\multirow{6}{*}{\makecell{\textbf{Qwen3-235B} \\ \textbf{-A22B-it(MoE)}}}
& Random
& \multirow{6}{*}{9.1}
& 9.1 & 9.3 & 9.4 & 9.6 & 9.9
& \multirow{6}{*}{18.6}
& 18.7 & 18.8 & 19.0 & 19.6 & 19.9 \\

& SN
&
& 9.3 & 10.9 & 11.4 & 12.2 & 14.6
&
& 19.1 & 20.8 & 22.7 & 25.1 & 27.3 \\

& GMT
&
& 9.8 & 11.0 & 12.5 & 14.7 & 16.2
&
& 19.6 & 21.3 & 24.9 & 26.5 & 29.7 \\

& Wanda
&
& 10.1 & 11.4 & 13.5 & 15.8 & 18.0
&
& 20.6 & 22.4 & 25.7 & 28.8 & 32.7 \\

& SNIP
&
& 11.3 & 13.0 & 15.4 & 17.8 & 20.3
&
& 22.1 & 24.6 & 27.8 & 31.5 & 35.2 \\

\rowcolor{Gray}
\cellcolor{white}
& \textbf{ESI}
&
& \textbf{27.6} & \textbf{30.2} & \textbf{33.1} & \textbf{37.9} & \textbf{41.1}
&
& \textbf{40.6} & \textbf{43.7} & \textbf{47.1} & \textbf{49.2} & \textbf{53.8} \\
\bottomrule
\end{tabular}

\caption{Perturbation analysis on additional models not included in the main experiments.
We report ASR (\%) on HarmBench and WildJailbreak under different parameter perturbation ratios.}
\label{tab:appendix_all_extra_models}
\end{table*}

\subsection{Experimental Setup}
To comprehensively verify the scalability and robustness of the proposed ESI framework across a broader spectrum of model sizes and architectural designs, we extend our perturbation-based sensitivity analysis to three additional large-scale LLMs. 
In the category of dense architectures, we select Qwen2.5-72B-it as a representative baseline to validate the efficacy of ESI on high-parameter dense structures. 
Furthermore, to rigorously assess the applicability of our method MoE architectures, we incorporate both Mixtral-8$\times$7B-it-v0.1 and the massive Qwen3-235B-A22B-it as representative models.

\subsection{Implementation of top-$k$\% Selection}
Directly identifying the global top-$k$\% parameters via ESI scores ($|\sigma(\theta_i)\nabla_{\theta_i}\mathcal{S}(\theta)|$) presents significant memory challenges for large-scale LLMs, such as Llama-3-70B-it. A standard global sort requires simultaneously storing gradients for all parameters, inevitably causing Out-Of-Memory (OOM) errors on typical GPUs. To address this, inspired by prior parameter-efficient selection methods~\cite{xie2024rtop,zhang2023parallel,li2024radik}, we propose a memory-efficient Distributed Threshold-based Selection (DTS) strategy. This approach circumvents full-model storage by processing parameters in three logical stages:

\paragraph{Stage 1: Threshold Estimation.} 
Rather than sorting the entire parameter space, we first estimate a rough cutoff threshold. 
We randomly sample a small fraction (e.g., 1\%) of parameters from each layer to construct a representative subset~\cite{greenwald2001space}. 
Based on this subset, we calculate a provisional threshold $\tau_{est}$ targeting the top-$(\lambda k)$\% percentile. 
We introduce a relaxation coefficient $\lambda$ (set to 1.5) to slightly lower the threshold, ensuring that the true top-$k$\% parameters are included despite potential sampling variance.

\paragraph{Stage 2: Layer-wise Filtering.} 
Using the estimated $\tau_{est}$, we process the model sequentially, layer by layer. 
For each layer $l$, we compute the ESI scores and immediately filter out parameters below the threshold:
\setcounter{equation}{0}
\begin{equation}
    M_{l} = \{ (\theta_i, s_i) \mid \theta_i \in \text{Layer}_l, s_i > \tau_{est} \}
\end{equation}
Only the candidate parameters in $M_l$ are transferred to CPU memory, after which the dense GPU tensors are instantly released~\cite{rajbhandari2020zero}. 
This strategy strictly bounds peak GPU memory usage to the size of a single layer rather than the entire model.

\paragraph{Stage 3: Global Exact Selection.} 
Finally, we aggregate the candidate sets $\{M_l\}$ from all layers on the CPU. 
Since the relaxation coefficient $\lambda$ yields a candidate pool slightly larger than the target $k$\%, we perform an exact sort on this reduced subset to identify the final global safety-critical parameters. 
This method reduces space complexity from $O(N)$ to approximately $O(\lambda k + \max(|\text{Layer}_l|))$, enabling the analysis of models exceeding 70 billion parameters on a single GPU.

\subsection{Baseline Descriptions}
We compare ESI against several established parameter importance metrics:
\begin{itemize}[leftmargin=*]
    \item \textbf{Random:} For random selection, parameters are sampled uniformly at random from the entire model
parameter space without relying on any gradient-based or task-specific signals.
After sampling, the selected parameters are subjected to the same perturbation procedure
as in other settings.
We consider selection ratios of 0.1\%, 0.5\%, 1\%, 3\%, and 5\% in our experiments.

    \item \textbf{SNIP~\citep{lee2019snip}:} SNIP utilizes a gradient-based sensitivity metric to identify critical model weights
by estimating the first-order Taylor approximation of the loss change when individual
parameters are zeroed~\cite{lee2019snip}.
The core idea of SNIP lies in its importance score, defined as $I(W) = \mathbb{E}_{x \sim D} \left| W \odot \nabla_W \mathcal{L}(x) \right|$ where $\mathcal{L}(x)$ denotes the conditional negative log-likelihood of the model
generating a target safe response.
In our study, we apply SNIP to compute importance scores for all model parameters
using the prompts sampled from the AdvBench dataset, aiming to localize safety-critical regions of the LLM~\cite{advbench}.
Based on the resulting importance scores, parameters are ranked and the top-$k$\%
neurons are selected under different selection ratios.
Specifically, we consider selection ratios of 0.1\%, 0.5\%, 1\%, 3\%, and 5\%,
and investigate how perturbations to these high-scoring parameters affect model safety.

    \item \textbf{Wanda~\citep{wanda}:} Wanda identifies influential parameters by approximating an output-preserving sparsification objective.
Given a calibration dataset, all input activations corresponding to a weight matrix $W$ are collected into $X_{in} \in \mathbb{R}^{d_{in} \times n}$.
The goal is to apply an element-wise binary mask $M \in \{0,1\}^{d_{out} \times d_{in}}$ to $W$ such that the Frobenius norm of the resulting output change~\cite{frantar2023sparsegpt}, measured as the difference between $W X_{in}$ and $(M \odot W) X_{in}$, is minimized.
Following Wanda, this objective is approximately solved by assigning an importance score to each weight entry, defined as the element-wise product between the absolute value of the weight matrix and the activation strength.
Concretely, the importance score is computed as $I(W) = |W| \odot (\mathbf{1} \cdot \|X_{in}\|_2^{\top})$, where $\mathbf{1} \in \mathbb{R}^{d_{out}}$ denotes an all-one vector and $\|X_{in}\|_2 \in \mathbb{R}^{d_{in}}$ represents the row-wise $L^2$ norm of the input activations.
This metric assigns higher importance to weights that are both large in magnitude and associated with strong activations, and pruning weight entries with smaller scores approximately minimizes the induced change in model outputs.
In our setting, we compute Wanda scores using the prompts sampled from the AdvBench dataset.
As we are only interested in measuring the contribution of each weight entry to the model’s generated responses, we mask out prompt activations and retain only response activations in $X_{in}$.
We then evaluate the model behavior by intervening on the top $0.1\%$, $0.5\%$, $1\%$, $3\%$, and $5\%$ of neurons ranked by the Wanda importance score.

    \item \textbf{GMT~\citep{GMT}:}  Gradient-Mask Tuning (GMT) is an in-training parameter selection method that selectively updates the most critical model parameters based on task-specific gradient information. The core of GMT lies in utilizing the absolute magnitude of accumulated gradients as a fine-grained saliency measure, defined as $s_{ij}=|\nabla_{\theta_{ij}}\mathcal{L}(\Theta;\mathcal{D})|$, to determine which weights exert the most substantial influence on the loss function~\cite{fu2023effectiveness,hui2025hft}. During the training process, a binary mask is applied to filter out gradients with small absolute values, ensuring that only those falling within a pre-defined top percentile $k$ are utilized for parameter updates. In our experimental setup, we apply the GMT approach to the prompts sampled from the AdvBench dataset to locate safety-relevant parameters. To evaluate the localization across different granularities, we configured the update percentile $k$ to target the top 0.1\%, 0.5\%, 1\%, 3\%, and 5\% of the total parameters, systematically observing how these salient updates contribute to the model's safety alignment.
    \item \textbf{SN~\citep{zhao2025understanding}:} The SN method identifies safety-specific neurons, defined as individual rows or columns of parameter matrices, that are consistently instrumental in processing and defending against harmful queries. The core importance of a neuron is quantified by the $L_2$ norm difference in intermediate representations upon its deactivation, expressed as $\|h_{\backslash N_{i}^{(l)},i}(x)-h_{i}(x)\|_{2}$. Unlike global ranking methods, this approach defines a safety subnetwork $\mathcal{N}_{safe}$ by extracting neurons that remain consistently activated across a diverse corpus of harmful queries. In our experimental setup using the prompts sampled from the AdvBench dataset, we localized safety parameters by specifically adjusting the number of top-activated neurons in both Feed-Forward (FFN) and Attention (ATTN) layers. By tailoring these layer-specific top-K counts, for example by selecting the top 1,200 parameters from FFN modules and the top 200 parameters from ATTN, corresponding to approximately 1\% of the model, we systematically approximate total parameter selection ratios of 0.1\%, 0.5\%, 1\%, 3\%, and 5\% to evaluate the robustness of the localized safety regions.

\end{itemize}

\subsection{Extended Perturbation Analysis}
\label{subsec:ext_perturbation}

Table~\ref{tab:appendix_all_extra_models} extends our perturbation analysis to diverse architectures, including MoE and ultra-large models. The results indicate that perturbing parameters identified by ESI leads to substantially higher ASR compared to all baselines. For instance, on Mixtral-8x7B-it, perturbing 5\% of parameters identified by ESI increases HarmBench ASR from 20.1 to 75.1, whereas the strongest baseline (SNIP) only reaches 59.6. Meanwhile, random perturbation results in negligible ASR changes across all settings, confirming that the safety degradation stems from our precise identification rather than random noise. We also observe that while the ultra-large model (Qwen3-235B-it) exhibits greater robustness, ESI still consistently maintains a clear margin over other methods. These results further verify the robustness and generalizability of ESI across different model scales and architectures.

\begin{table}[t]
\centering
\renewcommand{\arraystretch}{0.9}
\setlength{\tabcolsep}{4pt}
\footnotesize

\begin{tabular}{@{} l @{\hspace{4mm}} l cccc @{}}
\toprule
\textbf{Model} & \textbf{Judge Model} & \textbf{Base} & \textbf{0.1} & \textbf{0.5} & \textbf{1.0} \\
\midrule

\multirow{2}{*}{\makecell[l]{Qwen2.5\\14B-base}}
& GPT-Fuzz    & 55.1 & 73.1 & 75.9 & 78.3 \\
& Llama-Guard & 55.1 & 73.5 & 76.8 & 78.5 \\
\midrule

\multirow{2}{*}{\makecell[l]{Llama3\\8B-it}}
& GPT-Fuzz    & 15.3 & 42.1 & 55.8 & 59.3 \\
& Llama-Guard & 15.3 & 42.4 & 56.2 & 59.1 \\
\midrule

\multirow{2}{*}{\makecell[l]{Llama3\\70B-it}}
& GPT-Fuzz    & 16.2 & 44.0 & 49.3 & 56.8 \\
& Llama-Guard & 16.2 & 44.2 & 49.1 & 56.3 \\
\midrule

\multirow{2}{*}{\makecell[l]{Qwen3\\30B-A3B-it}}
& GPT-Fuzz    & 3.2  & 17.4 & 21.6 & 24.0 \\
& Llama-Guard & 3.2  & 17.5 & 22.1 & 24.6 \\
\midrule

\multirow{2}{*}{\makecell[l]{Qwen2.5\\14B-it}}
& GPT-Fuzz    & 13.0 & 26.2 & 31.0 & 36.7 \\
& Llama-Guard & 13.0 & 26.3 & 30.7 & 37.1 \\
\midrule

\multirow{2}{*}{\makecell[l]{Mixtral\\8$\times$7B-it}}
& GPT-Fuzz    & 20.1 & 51.2 & 56.8 & 64.1 \\
& Llama-Guard & 20.1 & 53.5 & 58.3 & 66.8 \\
\midrule

\multirow{2}{*}{\makecell[l]{Qwen3\\235B-A22B-it}}
& GPT-Fuzz    & 9.1  & 26.4 & 28.7 & 32.6 \\
& Llama-Guard & 9.1  & 27.6 & 30.2 & 33.1 \\

\bottomrule
\end{tabular}

\caption{
HarmBench ASR (\%) under parameter perturbation using ESI derived from GPT-Fuzz and Llama-Guard judge models.
Columns correspond to perturbation ratios (in \%).
}
\label{tab:esi_narrow_base_01_05_1}
\end{table}

\subsection{Effectiveness of Different Judge Models}

\label{ref_judgemodel_strategy}

To verify the robustness of the ESI framework, we evaluate the effectiveness of our judge-guided differentiable estimation across various LLMs. Specifically, we implement the estimation method using two distinct judge models: Llama-Guard and GPTFuzz.

As shown in Table \ref{tab:esi_narrow_base_01_05_1}, our framework exhibits strong robustness to the choice of the judge model. When comparing the results between Llama-Guard and GPTFuzz, we observe negligible differences in Attack Success Rate (ASR) across all evaluated models. Whether evaluating instruction-tuned models (e.g., Llama3-8B-it) or under-aligned base models (e.g., Qwen2.5-14B-base), the performance gap between the two judges typically remains within 0.5

This high consistency is crucial, as it suggests that the ESI metric successfully captures the intrinsic safety mechanisms of the target LLM itself, rather than overfitting to the specific preferences or labeling biases of an external evaluator. Overall, these results demonstrate that the proposed judge-guided estimation is a stable, reliable, and versatile methodology for identifying safety-critical parameters at various alignment stages.

\section{Additional Experiments on Safety Enhancement Tuning (SET)}
\label{sec:appendix_set_experiments}

In this section, we provide a detailed analysis of the implementation settings, baseline comparisons, and the impact of SET on general model capabilities compared to full parameter fine-tuning.

\subsection{Implementation Details}
We perform all fine-tuning using the AdamW optimizer with a learning rate of $2 \times 10^{-5}$, a cosine scheduler, and a 0.03 warmup ratio. To ensure memory efficiency, we enable gradient checkpointing and set weight decay to 0.001. We utilize 800 training samples with a per-device batch size of 1 and 8 gradient accumulation steps; this results in an effective batch size of 8, which aligns with our lightweight intervention strategy. 
Detailed hyperparameters are listed in Table~\ref{tab:hyperparameters of set}.

\begin{table}[h]
    \centering
    \setlength{\tabcolsep}{10pt} 
    \begin{tabular}{l|c}
    \toprule
    \textbf{Hyperparameter} & \textbf{Value} \\
    \midrule
    Optimizer & AdamW \\
    Learning Rate & $2 \times 10^{-5}$ \\
    LR Scheduler & Cosine \\
    Warmup Ratio & 0.03 \\
    Weight Decay & 0.001 \\
    \midrule
    Total Samples & 800 \\
    Per-Device Batch Size & 1 \\
    Gradient Accumulation Steps & 8 \\
    \bottomrule
    \end{tabular}
    \caption{Fine-tuning hyperparameters and implementation details of SET.}
    \label{tab:hyperparameters of set}
\end{table}

\subsection{Baselines}
To validate the effectiveness of our proposed strategy, we compare SET against the following fine-tuning methods. Note that for a fair comparison, all parameter selection baselines are restricted to the same update budget ($1\%$).

\begin{itemize}[leftmargin=*]
    \item \textbf{Random Selection:} Updates a random $1\%$ subset of parameters. This serves as a control baseline to verify the necessity of our targeted identification strategy.
    \item \textbf{SN-Tune:} Fine-tunes the top-$1$\% critical parameters identified by the Safety Neurons metric. This represents a baseline based on neuron-level safety analysis.
    \item \textbf{LoRA:} The standard parameter-efficient fine-tuning method implemented via Low-Rank Adaptation. We configure it with a rank of 64 and a learning rate of $5 \times 10^{-5}$ to serve as a general baseline.
    \item \textbf{SafeLoRA:} A safety-aware variant that constrains parameter updates to a safety-aligned subspace. We construct this subspace using the weight difference between the official instruction-tuned model and the base model. For implementation, we strictly follow the original setting with a rank of 64 and a learning rate of $5 \times 10^{-5}$.
\end{itemize}

\subsection{Impact on General Capabilities and Comparison with Full Fine-tuning}
\label{set_general}

\paragraph{Comparison with Full Fine-tuning.}
To validate the effectiveness of SET, we compare it against full parameter fine-tuning (FullFT). While FullFT theoretically maximizes safety by updating all model parameters, our results demonstrate that SET achieves nearly identical performance. As shown in Table~\ref{tab:SET vs fullft}, on the Qwen2.5-7B model trained with CB-Safety, FullFT reduces the ASR on HarmBench from 72.4 to 6.0. SET reaches a comparable ASR of 7.2, resulting in a negligible difference of only 1.2 points. We observe a similar trend on Llama3-8B trained with R1-Safety, where the performance gap on WildJailbreak is merely 1.5 points between the two methods. This result is significant given the computational difference. While FullFT requires updating 100\% of the parameters, SET achieves these safety gains by updating only the top 1\% critical weights. This confirms that SET is highly efficient, providing the safety benefits of full fine-tuning with significantly fewer resources.

\begin{table}[ht]
\centering
\resizebox{\columnwidth}{!}{
\begin{tabular}{ll cc | cc}
\toprule
\multirow{2}{*}{\textbf{Model}} & \multirow{2}{*}{\textbf{Method}} 
& \multicolumn{2}{c|}{\textbf{R1-Safety}} 
& \multicolumn{2}{c}{\textbf{CB-Safety}}\\
\cmidrule(lr){3-4} \cmidrule(lr){5-6}
 & & HB $\downarrow$ & WJ $\downarrow$ & HB $\downarrow$ & WJ $\downarrow$ \\ 
\midrule

\multirow{3}{*}{\makecell[l]{\textbf{Qwen2.5}\\\textbf{-7B-base}}} 
 & Base   
 & 72.4 & 77.2 & 72.4 & 77.2  \\

 & FullFT
 & 18.9 {\small \color{BlueDelta} $\Delta$53.5$\downarrow$}
 & 25.0 {\small \color{BlueDelta} $\Delta$52.2$\downarrow$}
 & 6.0  {\small \color{BlueDelta} $\Delta$66.4$\downarrow$}
 & 18.7 {\small \color{BlueDelta} $\Delta$58.5$\downarrow$} \\

 & \cellcolor{Gray}\textbf{SET}
 & \cellcolor{Gray}20.3 {\small \color{BlueDelta} $\Delta$52.1$\downarrow$}
 & \cellcolor{Gray}26.5 {\small \color{BlueDelta} $\Delta$50.7$\downarrow$}
 & \cellcolor{Gray}7.2  {\small \color{BlueDelta} $\Delta$65.2$\downarrow$}
 & \cellcolor{Gray}20.1 {\small \color{BlueDelta} $\Delta$57.1$\downarrow$} \\

\midrule

\multirow{3}{*}{\makecell[l]{\textbf{Qwen2.5}\\\textbf{-14B-base}}} 
 & Base   
 & 55.1 & 67.6 & 55.1 & 67.6 \\

 & FullFT
 & 5.8  {\small \color{BlueDelta} $\Delta$49.3$\downarrow$}
 & 13.2 {\small \color{BlueDelta} $\Delta$54.4$\downarrow$}
 & 2.9  {\small \color{BlueDelta} $\Delta$52.2$\downarrow$}
 & 8.9  {\small \color{BlueDelta} $\Delta$58.7$\downarrow$} \\

 & \cellcolor{Gray}\textbf{SET}
 & \cellcolor{Gray}7.4  {\small \color{BlueDelta} $\Delta$47.7$\downarrow$}
 & \cellcolor{Gray}4.7 {\small \color{BlueDelta} $\Delta$52.9$\downarrow$}
 & \cellcolor{Gray}4.1  {\small \color{BlueDelta} $\Delta$51.0$\downarrow$}
 & \cellcolor{Gray}10.1 {\small \color{BlueDelta} $\Delta$57.5$\downarrow$} \\

\midrule

\multirow{3}{*}{\makecell[l]{\textbf{Llama3}\\\textbf{-8B-base}}} 
 & Base   
 & 41.2 & 62.5 & 41.2 & 62.5 \\

 & FullFT
 & 5.9  {\small \color{BlueDelta} $\Delta$35.3$\downarrow$}
 & 17.6 {\small \color{BlueDelta} $\Delta$44.9$\downarrow$}
 & 4.0  {\small \color{BlueDelta} $\Delta$37.2$\downarrow$}
 & 12.9 {\small \color{BlueDelta} $\Delta$49.6$\downarrow$} \\

 & \cellcolor{Gray}SET
 & \cellcolor{Gray}7.4  {\small \color{BlueDelta} $\Delta$33.8$\downarrow$}
 & \cellcolor{Gray}19.1 {\small \color{BlueDelta} $\Delta$43.4$\downarrow$}
 & \cellcolor{Gray}5.2  {\small \color{BlueDelta} $\Delta$36.0$\downarrow$}
 & \cellcolor{Gray}14.3 {\small \color{BlueDelta} $\Delta$48.2$\downarrow$} \\

\bottomrule
\end{tabular}%
}
\caption{
ASR comparison on HarmBench (HB) and WildJailbreak (WJ) under full fine-tuning (FullFT) and selective fine-tuning (SET).
Models are fine-tuned using R1-Safety and CB-Safety datasets.
FullFT achieves slightly lower ASR, while SET attains comparable safety performance with substantially fewer updated parameters.
}
\label{tab:SET vs fullft}
\end{table}

\begin{figure}[t]
  \centering
  \includegraphics[width=\linewidth]{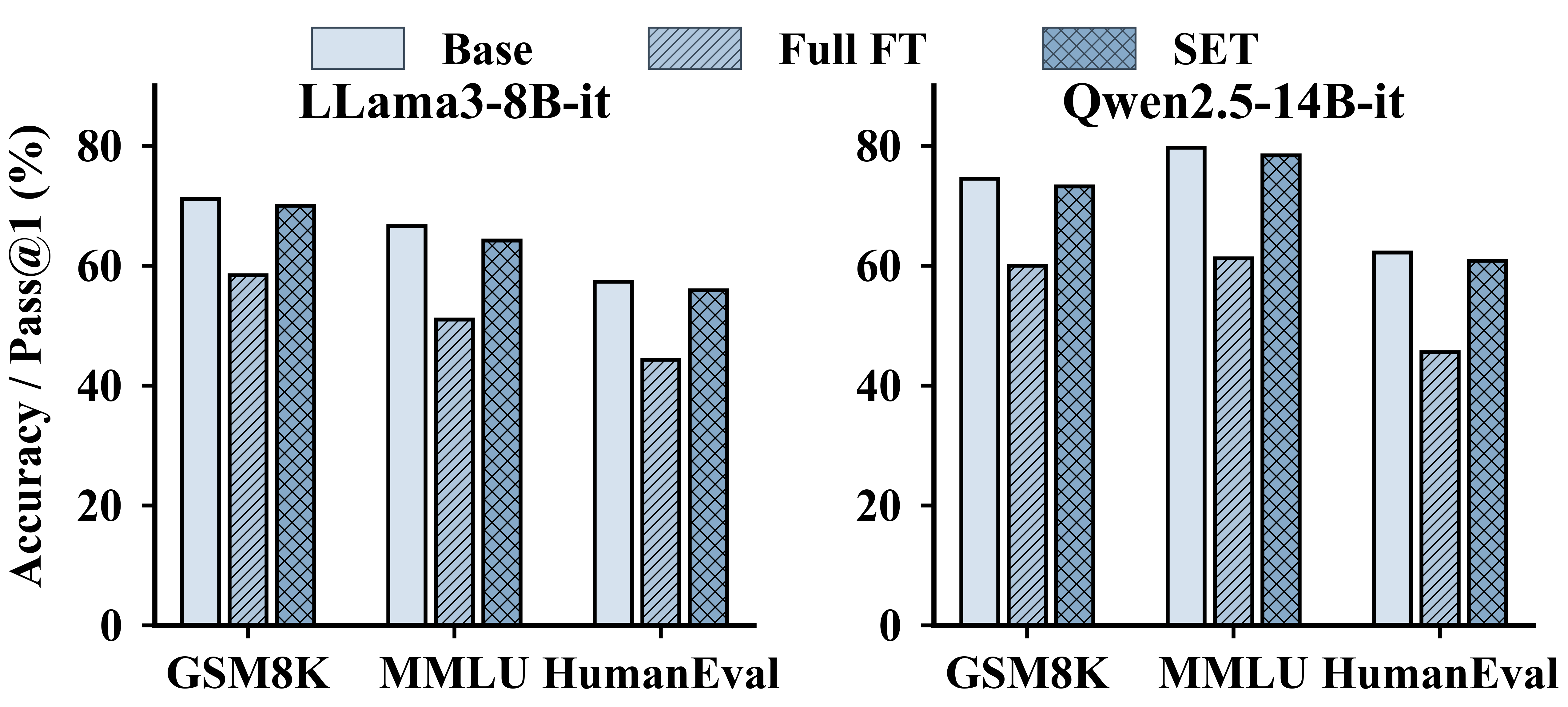}
  \caption{
General capability comparison of Base, Full Fine-Tuning (Full FT), and SET on Llama3-8B-it and Qwen2.5-14B-it across GSM8K, MMLU, and HumanEval.
}

  \label{fig:utility_single_col-it}
\end{figure}

\paragraph{Preservation of General Capabilities.}
In addition to safety, we must evaluate whether our method harms the model's general capabilities. We compared SET against Full Fine-Tuning (Full FT) and the Base models using GSM8K for reasoning, MMLU for knowledge, and HumanEval for coding. As shown in Figure \ref{fig:utility_single_col-it}, Full FT consistently degrades performance. For example, Llama3-8B-it showed a significant accuracy drop on GSM8K after Full FT, which indicates that updating all parameters causes the model to forget its reasoning skills. In contrast, SET maintains utility scores nearly identical to the Base model. Since SET only updates the top-1\% of parameters, it improves safety without sacrificing the model's core abilities.

\begin{figure*}[t] 
    \centering
    \includegraphics[width=0.85\linewidth]{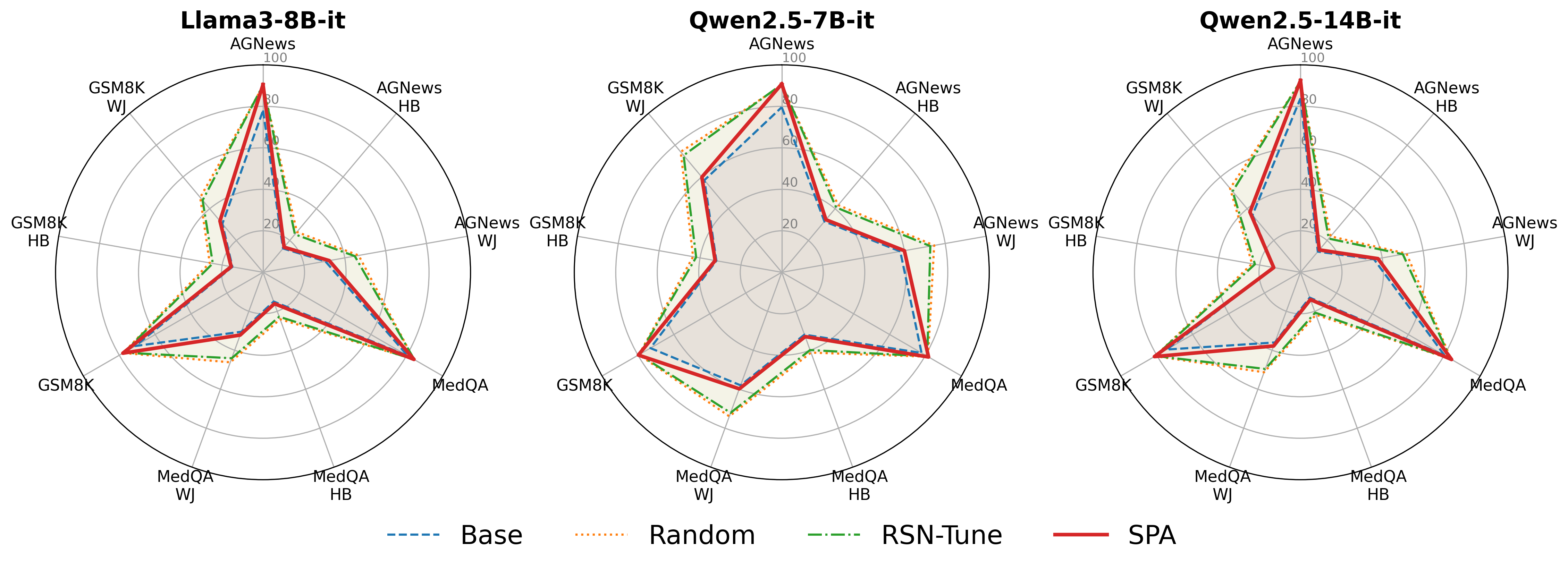}
    
    \caption{Radar charts illustrating the trade-off between safety and utility across three LLM architectures. We compare our method (SPA) against Base, Random, and RSN-Tune settings. The axes represent utility metrics (Accuracy/Score on AGNews, MedQA, GSM8K) and safety risks (Attack Success Rate on HarmBench and WildJailbreak). Note that for utility metrics, higher is better (outer edge), while for safety metrics (ASR), lower is better (inner center).}
    \label{fig:Radar_Chart_Final_Version}
\end{figure*}

\section{Experimental Details for Safety Preserving Adaptation (SPA)}
\label{app:spa_details}

This section provides the experimental setup for SPA. To ensure reproducibility, we detail the specific hyperparameter settings and baseline methods used in our evaluation. Additionally, we visualize the comprehensive performance trade-off between safety and utility using a radar chart in Figure~\ref{fig:Radar_Chart_Final_Version}.

\subsection{Implementation Details}
\label{app:spa_implementation}

We configure the learning rate at $2\times 10^{-5}$ and employ a cosine decay scheduler. To manage memory efficiency, we use a micro-batch size of 1 with gradient accumulation performed every 4 steps. Regarding the training data, we generally utilize 4,000 samples for each task and train for 1 epoch. The exception is MedicalQA, where we sample 2,000 instances and train for 2 epochs.



\subsection{Baselines}
To validate the effectiveness of SPA, we compare it against the following baselines under the identical parameter update budget (10\%):
\begin{itemize}[leftmargin=*]
    \item \textbf{Random Selection:} A straightforward baseline where 10\% of the model parameters are randomly selected for fine-tuning, while the remaining 90\% are frozen. This serves as a control group to demonstrate the necessity of targeted parameter selection.
    \item \textbf{RSN-Tune:} RSN-Tune is a structured baseline that fine-tunes a subset of safety-related parameters that do not overlap with foundation parameters, while freezing all remaining parameters. By explicitly separating safety-critical parameters from those essential for general task performance, this baseline is designed to evaluate whether avoiding such overlap can improve robustness against safety degradation during downstream fine-tuning.
\end{itemize}

\section{Ethical Considerations}
In this work, we propose the ESI framework to mechanistically understand and control LLM safety. Our research involves the use of benchmark datasets containing harmful prompts, such as AdvBench, HarmBench, and WildJailbreak. We emphasize that these datasets are used strictly for evaluating the effectiveness of our safety enhancement (SET) and preservation (SPA) methods in a controlled research setting.

\section{Use of AI Assistants}
We used AI assistants solely for editorial refinements, such as grammar and spelling checks, to enhance the clarity of the manuscript. All original research ideas, technical content, and experimental analyses were produced independently by the authors.

\section{Artifact Licenses and Intended Use}
All models and datasets used in this research are publicly available and utilized in accordance with their respective open-source licenses. We strictly adhere to the terms and conditions specified by the original creators regarding the use and distribution of these artifacts. Our utilization of all artifacts is strictly limited to academic research and safety analysis. This usage is fully consistent with the intended purposes and original access conditions defined by the developers of these models and datasets.

\end{document}